\documentclass[lettersize,journal]{IEEEtran}
\usepackage{amsmath,amsfonts}
\usepackage{algorithm}
\usepackage{array}
\usepackage{textcomp}
\usepackage{stfloats}
\usepackage{url}
\usepackage{verbatim}
\usepackage{graphicx}
\usepackage{cite}
\usepackage{amsthm}
\usepackage{subfigure}
\newtheorem{theorem}{Theorem}
\newtheorem{corollary}{Corollary}
\usepackage{algpseudocode}

\begin{document}

\title{Space-Time Coded RIS-Assisted Wireless Systems with Practical Reflection Models: Error Rate Analysis and Negative Moment-Based Optimization with Saddle Point Approximation}

\author{Tayfun Yilmaz, Haci Ilhan and İbrahim Hokelek
\thanks{T. Yilmaz is with the Department of Aviation Electrics and Electronics, Kocaeli University, Turkiye, and also with the Department of Electronics and Communication Engineering, Yildiz Technical University, 34220 Esenler, Istanbul, Turkiye (e-mail: tayfun.yilmaz@kocaeli.edu.tr).}
\thanks{H. ILHAN is with the Electronics and Communication Engineering, Yildiz Technical University, Turkiye (e-mail: ilhanh@yildiz.edu.tr).}
\thanks{I. Hokelek is with the Communications and Signal Processing Research (HİSAR) Lab., T{\"{U}}B{\.{I}}TAK B{\.{I}}LGEM, Turkiye 
(e-mail: ibrahim.hokelek@tubitak.gov.tr).}
}

\markboth{Possible IEEE Journal,~Vol.~XX, No.~X, August~2025}%
{Shell \MakeLowercase{\textit{et al.}}: A Sample Article Using IEEEtran.cls for IEEE Journals}


\maketitle

\begin{abstract}
Reconfigurable Intelligent Surface (RIS)-assisted communication has recently attracted significant attention for enhancing wireless performance in challenging environments, making accurate error analysis under practical hardware constraints crucial for future multi-antenna systems. This paper presents a theoretical framework for symbol error rate (SER) analysis of RIS-assisted multiple-antenna systems employing orthogonal space-time block codes (OSTBC) under practical reflection models with amplitude-dependent and quantized phase responses. 
By exploiting the Gramian structure of the cascaded channel \(\mathbf{f} \), we derive exact moment-generating function (MGF) expressions of the nonzero eigenvalue of \( \mathbf{f}^{\dagger} \mathbf{f} \) for small RIS sizes. For large-scale RIS deployments, where closed-form analysis becomes intractable, we employ Saddle Point Approximation (SPA) to approximate the eigenvalue distribution. Using these results, we derive unified SER expressions using exact and SPA-based MGF formulations, applicable to arbitrary RIS sizes, phase configuration, and both identical and non-identical amplitude responses. Extensive Monte Carlo simulations confirm the accuracy of the proposed SER expressions, demonstrating very close agreement for all configurations. 
In addition, by applying asymptotic SNR analysis on the SPA-based SER formulation, we mathematically establish that the coding gain is inversely proportional to the $N_t$-th negative moment of the SPA-approximated probability density function (PDF) corresponding to the nonzero eigenvalue of the cascaded RIS–receiver Gram matrix. 
This insight motivates a negative moment minimization algorithm that efficiently identifies hardware-constrained RIS phase configurations, achieving near-optimal SER performance with low complexity.
\end{abstract}


\maketitle

\section{INTRODUCTION}
\label{Sec_Intro}
\IEEEPARstart{R}{econfigurable} Intelligent Surface (RIS) technology has recently emerged as a promising solution for enhancing wireless communication by improving coverage, energy efficiency, information security, and throughput in next-generation systems such as massive MIMO networks~\cite{GongSurvey2021, DiRenzoTutorial2020, RIS-EFF-COV, RIS-SEC-RATE, RIS-ERG-MIMO}. These advantages stem from the ability of RIS to dynamically control the amplitude and phase of reflected electromagnetic waves, thereby enabling programmable radio environments~\cite{ZhouOptSurvey2023, SEFA}. For example, the integration of RIS into multiple-antenna systems meets extremely high throughput and reliability demands of 5G and 6G networks \cite{RIS-INDEX-MOD}. In such systems, orthogonal space-time block coding (OSTBC) schemes \cite{OSTBC-V.TAROKH} are expected to play a key role due to their ability to provide full spatial diversity while maintaining low decoding complexity and operating without requiring full channel state information (CSI) at the transmitter.

Several recent studies have explored the use of space-time block codes in RIS-assisted wireless systems \cite{RIS-MIMO-ASTBC, GSTBC-SM-RIS, STBC-DSTBC-RIS, OSTBC-RIS-SENSORS}. For instance, in \cite{RIS-MIMO-ASTBC}, Khaleel and Basar propose a novel transmitter architectures based on Alamouti-STBC and vertical Bell Laboratories Layered Space-Time (V-BLAST) schemes, where an RIS equipped with an RF generator is deployed at the transmitter side. In \cite{GSTBC-SM-RIS}, a generalized STBC-based spatial modulation (GSTBC-SM) scheme is introduced to support various RIS configurations, where the RIS operates as a transmitter, a passive relay, or an active relay. However, this work does not provide closed-form expressions for instantaneous SNR statistics, and the performance analysis is limited to numerical upper bounds on the average bit error rate. Furthermore, the results in \cite{GSTBC-SM-RIS} are valid primarily for systems with a large number of reflecting elements. In \cite{STBC-DSTBC-RIS}, closed-form and asymptotic SER expressions are derived for STBC and differential STBC schemes supported by RIS \cite{STBC-DSTBC-RIS}. As the RIS is used as a transmitter, the results do not include SER expressions for the cascaded fading condition. In the context of RIS-assisted communication systems employing OSTBC techniques, El-Hussein \textit{et al.} \cite{OSTBC-RIS-SENSORS} propose an algorithm to evaluate the error performance of the system. However, the study lacks validation through closed-form mathematical computations, as the results are entirely based on simulations. Finally, in \cite{ASTBC-RIS-SER}, the authors present closed-form SER and outage probability analyses for RIS-assisted Alamouti STBC; however, the study is limited to the Alamouti scheme and assumes identical amplitude responses across RIS elements. A common limitation in these space-time coded RIS-assisted studies is the assumption that the phase correction capabilities of RIS elements are ideal and identical, which does not reflect the practical constraints encountered in real-world implementations.

In addition to the STBC-based RIS-assisted multi-antenna systems, recent studies have also explored alternative RIS-MIMO architectures, considering both ideal assumptions and practically impaired hardware conditions. In \cite{RIS-MIMO-NEW1}, Sui \textit{et al.} derive closed-form expressions for the ergodic capacity of RIS-assisted MIMO systems, assuming that the signal reflection from any RIS element is ideal, i.e., without any power loss. Abbasi \textit{et al.} \cite{ abbasi2023rayleighrician, abbasi2024ergodic} develop ergodic capacity expressions under practical amplitude-phase reflection models, whereas Ramezani \textit{et al.} \cite{ramezani2024mse} consider joint optimization of RIS and precoding under fronthaul and quantization limits. Furthermore, Li \textit{et al.} \cite{li2024ergodicser} consider both ergodic rate and SER optimization in mmWave MIMO systems, while Chien \textit{et al.} \cite{beamforming2024ser} and Yang \textit{et al.} \cite{yang2024spatially} focus on beamforming and secrecy optimization under hardware impairments and imperfect CSI.

Despite these advances, the following critical gaps remain in the literature on RIS-assisted multiple-antenna systems: (i) Most of the existing studies do not provide unified closed-form expressions for SER under realistic RIS models involving phase-dependent and non-identical amplitude responses. While some works address SER numerically or through simulations, tractable analytical formulations remain absent. (ii) To the best of our knowledge, the only closed-form SER analysis for OSTBC considers only the Alamouti scheme under the ideal RIS element case, leaving general OSTBCs and realistic models unexplored (iii) Moreover, the asymptotic behavior of RIS-assisted systems with large numbers of elements has not been sufficiently studied using scalable mathematical tools such as the SPA, even though these tools can provide compact expressions valid across a wide range of configurations. (iv) Optimization techniques in prior works are largely based on ergodic rate or average SER metrics, which require numerically intensive computations or Monte Carlo simulations, making real-time or large-scale deployment challenging. (v) No existing study establishes a theoretical connection between coding gain and the negative moment of the SNR distribution, nor leverages this relationship for analytically tractable and hardware-constrained RIS configuration. (vi) Although several works have investigated space-time coding in RIS-assisted systems, to the best of our knowledge, no study has analyzed the error performance of such systems under phase-dependent amplitude response conditions.
\IEEEpubidadjcol

These limitations motivate the development of a scalable, interpretable, and unified mathematical framework for SER performance analysis of space-time coded RIS-assisted multiple-antenna systems, which is the focus of this work. Specifically, our contributions are summarized as follows:

\begin{itemize}
   \item By exploiting the Gramian structure of the cascaded fading channel, we derive exact MGF expressions for the nonzero eigenvalue of the cascaded fading channel vector under both identical and non-identical, phase-dependent amplitude response models. These MGF expressions enable the derivation of closed-form SER formulas for small RIS sizes and arbitrary phase configurations.
   \item For large-scale RIS scenarios, where the exact evaluation of the PDF corresponding to the nonzero eigenvalue of the cascaded RIS–receiver Gram matrix $\mathbf{\Phi^{\dagger} \mathbf{g}^{\dagger} \mathbf{g} \Phi}$ becomes analytically intractable, we approximate the PDF using Lindeberg-Feller central limit theorem (LCLT) and saddle point approximation (SPA). We demonstrate that SPA offers highly accurate approximations for arbitrary numbers of RIS elements. Based on the SPA-approximated PDF, we derive generalized MGF expressions and obtain scalable SER formulas valid for arbitrary RIS sizes and phase configurations.
   \item In addition, our SER analysis uniquely provides a unified framework for general OSTBC schemes under both ideal and practical RIS models.
   \item By applying an asymptotic SNR analysis to the SPA-based SER expression, we establish that the coding gain is inversely proportional to the $N_t$-th negative moment of the PDF corresponding to the nonzero eigenvalue of the RIS-receiver Gram matrix $\mathbf{\Phi^{\dagger} \mathbf{g}^{\dagger} \mathbf{g} \Phi}$. This insight enables a new RIS optimization paradigm that avoids the need for direct computation of SER or ergodic rate expressions.

   \item We propose a negative-moment-based optimization algorithm to configure RIS phase shifts. To demonstrate the applicability of the proposed $N_t$-th negative moment as an optimization objective, we implement a lightweight Greedy Search strategy~\cite{Int_to_Algo} based on element grouping. The resulting method yields hardware-constrained configurations and achieves near-optimal SER performance with significantly reduced complexity.
   \item All analytical expressions and the proposed optimization algorithm are validated through extensive Monte Carlo simulations, confirming near-perfect agreement for a wide range of RIS sizes, phase configurations, and amplitude response models. Overall, this study presents a unified framework for SER analysis and a generalized optimization strategy for space-time coded RIS-assisted multi-antenna systems under practical hardware constraints.
\end{itemize}

The remainder of the paper is organized as follows. Section~\ref{Sec_System_Model} introduces the proposed RIS-assisted multiple antenna system. Section~\ref{Sec_Per_An_Iden} presents the mathematical analysis for the case where the RIS elements have identical amplitude responses. Section~\ref{Sec_Per_An_Noniden} extends the analysis to the non-identical case and explores several analytical approaches. Section~\ref{sec_optimization} proposes a performance-enhancing optimization algorithm, which is developed based on key mathematical insights derived in the earlier sections. In Section~\ref{Sec_Num_Res}, the RIS path loss model is described, and the theoretical findings and the proposed algorithm are validated through detailed numerical simulations. Finally, Section~\ref{Sec_Conclusion} provides concluding remarks, highlighting the key limitations of the proposed approach, and presenting potential directions for future research.

$\textit{Notations}$: The superscripts $\left(\cdot\right)^{\textbf{T}}$ and $\left(\cdot\right)^{\dagger}$ denote the transpose and conjugate transpose operators, respectively.
The Euclidean (or 2-) norm of a complex vector $\mathbf{a}$ is denoted by $\|\mathbf{a}\|$, and its square is given by $\|\mathbf{a}\|^2 = \sum_i |a_i|^2$, where $|a_i|^2$ is the squared magnitude of each complex entry.
The function $f_a\left(\cdot\right)$ refers to the PDF of the random variable $a$, while $M_a\left(\cdot\right)$ denotes its MGF. The Laplace transform of a function $f(t)$ is represented as $\mathcal{L}\{f(t)\}$, and the expectation operator is denoted by $\mathbb{E}\left[\cdot\right]$. The operator $\mathrm{diag}(\mathbf{v})$ forms a diagonal matrix with the elements of vector $\mathbf{v}$. Additionally, matrices and vectors are represented using boldface letters, where uppercase letters (e.g., $\mathbf{A}$) denote matrices and lowercase letters (e.g., $\mathbf{a}$) denote vectors. $\Gamma\left(\cdot\right)$, $K_{a}\left(\cdot\right)$, $W_{a,b}\left(\cdot\right)$ and $U\left(a,b,c\right)$ respectively stand for the Gamma function~$\left[\citenum{Ryzhik},~ \text{Eq}.~(8.310.1)\right]$, modified Bessel-K function~$\left[\citenum{Abramowitz},~ \text{Eq}.~(9.6.2)\right]$, Whittaker's W-function~$\left[\citenum{Abramowitz},~ \text{Eq}.~(13.1.33)\right]$ and confluent hypergeometric Tricomi's function~$\left[\citenum{Abramowitz},~ \text{Eq}.~(13.2.5)\right]$.

\IEEEpubidadjcol
\begin{figure*}[t!]
    \centering {\includegraphics[width=6.0in, angle=0]{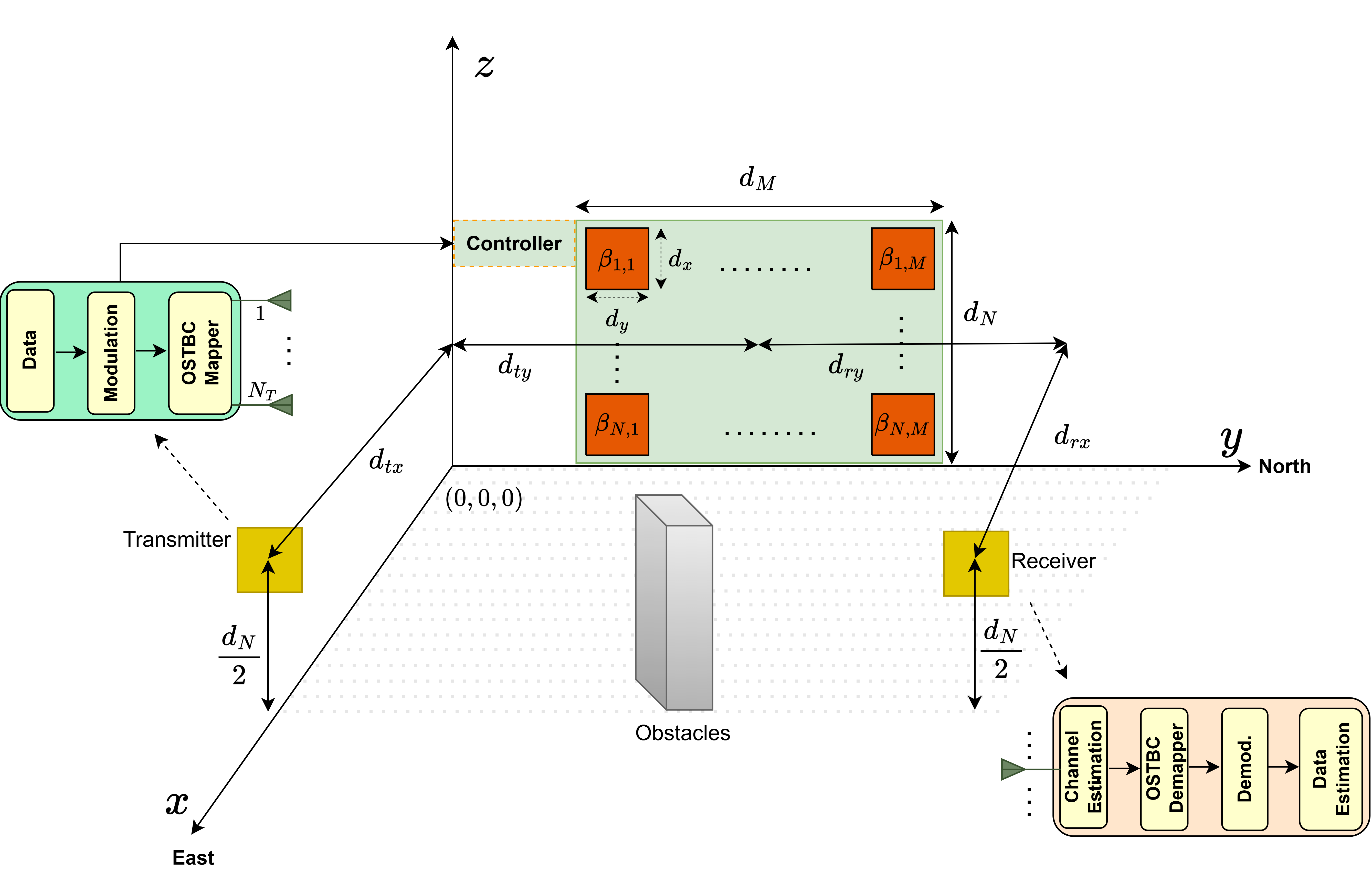}}
    \caption{Proposed space-time coded RIS-assisted multiple-antenna system model.}
    \label{fig.1}
\end{figure*}
\section{SYSTEM MODEL}
\label{Sec_System_Model}
In this section, we present a system model for a space-time coded RIS-assisted multiple-antenna system. As depicted in Fig.~\ref{fig.1}, a transmitter equipped with $N_t$ antennas and a receiver equipped with $N_r$ antennas communicate through an RIS composed of $N_{\text{RIS}} = M \times N$ reflecting elements. Due to the presence of obstacles, the direct link between the transmitter and receiver is blocked, and communication is established entirely through the RIS via a cascaded channel. In the proposed system, since OSTBC is employed as the diversity technique, it is assumed that the transmitter utilizes 2, 3, or 4 of its antennas for active transmission, while the receiver uses a single antenna at any given time instance to maintain compatibility with OSTBC decoding structures. To support this, it is further assumed that the subset of antennas actively used at the transmitter and receiver are selected dynamically using conventional antenna selection techniques with OSTBC techniques~\cite{Antenna_Selection}.

We consider a practical RIS model in which each element introduces not only a phase shift but also an amplitude response that depends on its configuration. The fading channels between the transmitter, RIS, and receiver are assumed to follow independent and identically distributed (i.i.d.) complex Gaussian distributions with zero mean and unit variance. This implicitly assumes that the minimum spacing between adjacent antenna elements at the transmitter and receiver, as well as the minimum spacing between adjacent RIS elements, is at least half of the carrier wavelength, thereby ensuring spatial independence of the fading coefficients.
\IEEEpubidadjcol

The receiver estimates the CSI through downlink cascaded channel state information reference signal (CSI-RS) symbols. The overall cascaded channel vector is modeled as
\begin{equation}
\label{Cascade-RIS-STBC}
\mathbf{f} = \mathbf{g} \mathbf{\Phi} \textbf{H},
\end{equation}
where $\textbf{H} \in \mathbb{C}^{N_{\text{RIS}} \times N_t}$ represents the fading channel matrix from the transmitter to the RIS, and $\mathbf{g} \in \mathbb{C}^{1 \times N_{\text{RIS}}}$ represents the fading channel vector from the RIS to the receiver. The $(n,m)$-th RIS element has a complex reflection coefficient denoted as $\beta_{n,m}(\phi_{n,m})e^{j\phi_{n,m}}$, where $\phi_{n,m} \in [0, 2\pi)$ is the phase shift and $\beta_{n,m}(\phi_{n,m}) \in [0,1]$ is the amplitude response. The phase adjustment matrix is defined as
\begin{equation}
\mathbf{\Phi} = \text{diag}\left(\beta_{1,1}(\phi_{1,1})e^{j\phi_{1,1}}, \dots, \beta_{N,M}(\phi_{N,M})e^{j\phi_{N,M}}\right).
\end{equation}
We consider a hardware-constrained RIS model, where the reflection amplitude depends on the applied phase shift \cite{RIS-AMP-RES}, following
\begin{equation}
    \label{Amp_Response}
    \beta_{n,m}(\phi_{n,m}) = (1 - \zeta_{\min}) \left( \frac{\sin(\phi_{n,m} - c)}{2} \right)^k + \zeta_{\min},
\end{equation}
where $\zeta_{\min} \geq 0$, $c \geq 0$, and $k \geq 0$ are constants defined by the surface impedance and circuit layout. This non-ideal model ($\beta_{n,m}(\phi_{n,m}) \neq 1$) captures the realistic behavior of reflecting elements and will later be used in both simulation and mathematical analysis.

The transmitter employs OSTBC to exploit spatial diversity.
The OSTBC generator matrix is denoted by $\textbf{G}_{N_t} = [x_{i,j}]_{T \times N_t}$ with code rate $R_{c} = K/T$\footnote[1]{Note that in this paper, we utilize the ${\textbf{G}_2}$ (Alamouti scheme) $\left[\citenum{OSTBC-V.TAROKH},\text{eq}.(3)\right]$, ${\textbf{G}_3}$ $\left[\citenum{OSTBC-V.TAROKH},\text{eq}.(4)\right]$, and ${\textbf{G}_4}$-OSTBC $\left[\citenum{OSTBC-V.TAROKH},\text{eq}.(5)\right]$ schemes. For more detailed information on these OSTBC code structures, please refer to \cite{OSTBC-V.TAROKH}.}. Each transmit symbol satisfies $\mathbb{E}\left[x_{i,j}x_{i,j}^{\dagger}\right] = P_t/N_t$. 
The received signal can then be expressed as
\begin{equation}
\label{Rx_Signs}
\mathbf{y} = \sqrt{P_{L}}\, \mathbf{G}_{N_t} \mathbf{f}^{\text{T}} + \mathbf{w},
\end{equation}
where $\mathbf{w} \in \mathbb{C}^{T \times 1}$ is an additive white Gaussian noise vector with zero mean and variance $\sigma_w^2$ per entry. The term $P_{L}$ represents the overall path-loss, which in our system accounts not only for distance attenuation but also for the elevation angle-dependent directional gain of the RIS, as well as the amplitude response of the reflecting elements. A detailed derivation of $P_{L}$ is provided in Section~\ref{Sec_Num_Res}.
\IEEEpubidadjcol

When OSTBC is used, the instantaneous SNR at the receiver is given by~\cite{OSTBC-V.TAROKH}
\begin{equation}
\label{SNR_Exp}
\rho = \frac{P_{L}}{R_{c}} \bar{\rho} \|\mathbf{f}\|^2,
\end{equation}
where $\bar{\rho} = \frac{P_t}{N_t \sigma_w^2}$ denotes the average transmit SNR. Since $\mathbf{f}$ is a complex-valued row vector, its squared norm $\|\mathbf{f}\|^2$ is equal to the unique nonzero eigenvalue of the Gram matrix $\mathbf{f}^\dagger \mathbf{f}$. Therefore, the SNR expression can be rewritten as
\begin{equation}
\label{SNR_Exp_2}
\rho = \frac{P_{L}}{R_{c}} \bar{\rho} Z,
\end{equation}
where $Z = z^2$ and $z$ is the singular value (or Euclidean norm) of $\mathbf{f}$. To determine the PDF of $Z$, we exploit the Gram matrix structure of $\mathbf{f}$. Specifically, $\mathbf{f}^\dagger \mathbf{f} \in \mathbb{C}^{N_t \times N_t}$ is a rank-one Hermitian matrix and has a single nonzero eigenvalue, which is equal to $Z$.
To determine the distribution of $Z = \|\mathbf{f}\|^2$, the instantaneous SNR expression can be reformulated as
\begin{equation}
\label{SNR_Gram}
\rho = \frac{P_{L}}{R_{c}} \bar{\rho} \, \mathbf{H}^{\dagger} \mathbf{\Phi}^{\dagger} \mathbf{g}^{\dagger} \mathbf{g} \mathbf{\Phi} \mathbf{H}.
\end{equation}

Through eigenvalue decomposition of the RIS–receiver Gram matrix $\mathbf{\Phi}^{\dagger} \mathbf{g}^{\dagger} \mathbf{g} \mathbf{\Phi}$, this expression can be rewritten as
\begin{equation}
\label{SNR_Eig}
\rho = \frac{P_{L}}{R_{c}} \bar{\rho} \, \mathbf{H}^{\dagger} \mathbf{\Theta} \mathbf{\Lambda} \mathbf{\Theta}^{\dagger} \mathbf{H},
\end{equation}
where
\begin{equation}
\mathbf{\Lambda} = \text{diag}\{ \lambda, \underbrace{0, \ldots, 0}_{N - 1} \},
\end{equation}
and $\lambda$ denotes the single nonzero eigenvalue of the matrix $\mathbf{\Phi}^{\dagger} \mathbf{g}^{\dagger} \mathbf{g} \mathbf{\Phi}$.
Since the distribution of $\mathbf{H}$ is invariant under left and right unitary transformations, \eqref{SNR_Eig} can be simplified as
\begin{equation}
\label{SNR_Final}
\rho = \frac{P_{L}}{R_{c}} \bar{\rho} \, \widetilde{\mathbf{H}}^{\dagger} \lambda \widetilde{\mathbf{H}},
\end{equation}
where $\widetilde{\mathbf{H}} \in \mathbb{C}^{1 \times N_t}$ is a statistically equivalent channel vector. Consequently, the SNR expression can be reformulated as
\begin{equation}
\label{SNR_Exp_Final}
\rho = \frac{\bar{\gamma}}{R_c} Z,
\end{equation}
where \( \bar{\gamma} = \bar{\rho} P_L \) denotes the average received SNR, and \( Z \) is the random variable defined as the single nonzero eigenvalue of \( \widetilde{\mathbf{H}}^{\dagger} \lambda \widetilde{\mathbf{H}} \). Thus, the characterization of \(Z\) reduces to the distribution of this quadratic form.

\section{ERROR RATE ANALYSIS FOR IDENTICAL RIS AMPLITUDE COEFFICIENTS}
\label{Sec_Per_An_Iden}
This section presents an error rate analysis for the scenario in which all RIS elements are assumed to have identical amplitude reflection coefficients. First, the statistical properties of the instantaneous SNR are derived by obtaining closed-form expressions for the PDF and MGF of the non-zero eigenvalue \( z \). These statistical tools are then employed to derive exact SER expressions for RIS-assisted space-time coded systems using \( M \)-ary phase shift keying (M-PSK) modulation. Additionally, an asymptotic high-SNR analysis is performed to characterize the achievable diversity and coding gains under this idealized amplitude configuration.

\subsection{PDF of \(Z\)}

The PDF of the instantaneous SNR is directly related to the PDF of $Z$ (i.e., $f_{Z}(a)$). To calculate $f_{Z}(a)$, the PDFs of $\lambda$, and $Z$ conditioned on $\lambda$, must first be derived. 
\begin{theorem}
\label{theo:pdf_lam_idn} 
Since \( \boldsymbol{\Phi} \) is a diagonal matrix and \( \boldsymbol{g} \) is a vector, the single non-zero eigenvalue of the matrix \( \boldsymbol{\Phi}^{\dagger} \mathbf{g}^{\dagger} \mathbf{g} \boldsymbol{\Phi} \) can be obtained by summing the diagonal elements of  \( \boldsymbol{\Phi}^{\dagger} \boldsymbol{\Phi} \mathbf{g}^{\dagger} \mathbf{g} \) as follows
 \begin{equation}
 \label{summ_exp}
     \lambda = \sum_{i=1}^{N_{\text{RIS}}}\beta^{2}_{i}(\phi_{i})\left|g_{i}\right|^2.
 \end{equation}
Here, \( \left|g_{i}\right|^2 \) follows an exponential distribution with unit mean and unit variance~\cite{shankar}. Moreover, when the amplitude responses of the RIS elements are identical, i.e., 
\begin{equation}
\beta_{1}^{2}(\phi_{1}) = \beta_{2}^{2}(\phi_{2}) = \dots = \beta_{N_{\text{RIS}}}^{2}(\phi_{N_{\text{RIS}}}) = \beta^{2}(\phi),
\end{equation}
the summation becomes a sum of independent and identically distributed exponential random variables. Consequently, the non-zero eigenvalue \( \lambda \) follows an Erlang distribution, expressed as
\begin{equation}
\lambda \sim \text{Erlang}\left(N_{\text{RIS}}, \frac{1}{\beta^{2}(\phi)}\right),
\end{equation}
where \( k = \frac{1}{\beta^{2}(\phi)} \) denotes the shape parameter~\cite{shankar}. Finally, the PDF of \( \lambda \) can be written as

\begin{equation}
\label{pdf_lambda}
f_{\lambda}(y) = \frac{ \left(\beta^{2}(\phi)\right)^{-N_{\text{RIS}}}}{\Gamma\left(N_{\text{RIS}}\right)}y^{N_{\text{RIS}} - 1} e^{-y/{\beta^{2}(\phi)}}, \quad y \geq 0.
\end{equation}
\end{theorem}

To calculate the PDF of \( Z \), the conditional PDF of \( Z \) given \( \lambda \) must be used. This expression is obtained by applying certain mathematical operations to~\cite[eq.~(95)]{Conditional_PDF_WISH}, and it can be expressed as 
\begin{equation} \label{pdf_z_beta} f_{Z|\lambda}(a) = \frac{a^{N_t-1}e^{-a/y}y^{-N_{t}}}{\Gamma\left(N_{t}\right)}, \quad a \geq 0. \end{equation} 

\begin{corollary}
\label{coro:pdf_z_idn}
The unconditional PDF of \( Z \) can be obtained by taking the expectation of the conditional PDF \( f_{Z|\lambda}(a) \) over \( \lambda \), as given by
\begin{equation}
\begin{aligned}
f_{Z}(a) &= \mathbb{E}_{\lambda}\left[ f_{Z|\lambda}(a)\right] = \frac{a^{N_{t}-1}\left(\beta^{2}(\phi)\right)^{-N_{\text{RIS}}}}{\Gamma\left(N_{t}\right)\Gamma\left(N_{\text{RIS}}\right)} \\ &\times\int_{0}^{\infty} y^{N_{\text{RIS}} - N_{t} - 1} \exp\left(-\frac{a}{y} - \frac{y}{\beta^{2}(\phi)}\right) dy.
\label{eqn:fz_app}
\end{aligned}
\end{equation}
As a result, the PDF of \( z \) can be obtained by solving the integral in~\eqref{eqn:fz_app} using~\cite[eq.~(3.471.9)]{Ryzhik}, yielding
\begin{equation}
\begin{aligned}
f_{Z}(a) &= \frac{2}{\Gamma\left(N_{t}\right)\Gamma\left(N_{\text{RIS}}\right)} a^{\frac{N_{\text{RIS}} + N_{t} - 2}{2}} \left(\beta(\phi)\right)^{-N_{\text{RIS}}-N_{t}} \\&\times K_{N_{\text{RIS}} - N_{t}}\left(\frac{2}{\beta(\phi)} \sqrt{a}\right).
\label{eqn:fz_final}
\end{aligned}
\end{equation}
\end{corollary}

\subsection{MGF of instantaneous SNR $\rho$}\label{subsec:mgf_rho} 
To derive analytical expressions of performance metrics such as SER and BER, the MGF of the instantaneous SNR must be expressed with a negative argument. In other words, the \textit{Laplace transform} of the instantaneous SNR distribution should be employed. The relationship between the Laplace transform and the MGF is given by
\begin{equation}
\mathcal{L}\left\{ f_{Z}(a) \right\} = M_{Z}(-s).
\end{equation}
\IEEEpubidadjcol

\begin{corollary}
\label{coro:mgf_rho_idn}
By applying the \textit{Laplace transform} on the PDF of \( Z \), and using~\eqref{SNR_Exp_Final} along with~\cite[eq.~(6.643.3)]{Ryzhik}, the MGF of the instantaneous SNR, \( M_{\rho}(-s) \), can be expressed in terms of the Whittaker-\( W \) function as follows:
\begin{equation}
\begin{aligned}
M_{\rho}(-s) &= \frac{\left(\beta(\phi)\right)^{1 - N_t-N_{\text{RIS}}} \exp\left(\frac{1}{2\mu(s)\beta^2(\phi)}\right)}{\left(\mu(s)\right)^{\frac{N_{\text{RIS}} + N_t - 1}{2}}} \\
& \times W_{\frac{1 - N_t - N_{\text{RIS}}}{2}, \frac{N_{\text{RIS}} - N_t}{2}}\left(\frac{1}{\mu(s)\beta^2(\phi)}\right),
\label{eqn:mgf_rho_whittaker}
\end{aligned}
\end{equation}
where \( \mu(s) = \frac{\bar{\gamma}}{R_c}s \). 
Finally, by employing the relationship between the Whittaker-\( W \) function and the confluent hypergeometric Tricomi's function~\cite[eq.~(13.1.33)]{Abramowitz}, and performing some algebraic manipulations, the expression in~\eqref{eqn:mgf_rho_whittaker} can be rewritten in a simplified form as
\begin{equation}
\begin{aligned}
M_{\rho}(-s) &= \mu(s)^{-N_{\text{RIS}}}\left(\beta(\phi)\right)^{-2N_{\text{RIS}}} \\ &\times U\left(N_{\text{RIS}}, N_{\text{RIS}} - N_t + 1, \frac{1}{\mu(s)\beta^2(\phi)}\right).
\label{mgf_z_Tricomi}
\end{aligned}
\end{equation}
\end{corollary}

Equation~\eqref{mgf_z_Tricomi} provides a simplified and exact expression for the MGF of the instantaneous SNR, under the assumption that the amplitude responses of all RIS elements are identical.
\subsection{SER calculation} The average SER expressions for the M-PSK modulation scheme can be expressed as follows\cite{SER-PSK-QAM},
\begin{equation}
\label{eqn:SER_PSK_Int}
    \bar{P}_{\text{M-PSK}} = \frac{1}{\pi}\int_{0}^{\frac{(\text{M}-1)\pi}{\text{M}}}M_{\rho}\left(-\frac{\alpha_{\text{PSK}}}{2\text{sin}^{2}\theta}\right)d\theta,
\end{equation}
where $\alpha_{\text{PSK}} = 2\sin^2\left(\frac{\pi}{\text{M}}\right)$, and $\text{M}$ represents the modulation order.
This MGF-based calculation offers a low-complexity approach for determining the SER of space-time coded RIS-assisted systems. Note that the accuracy of the calculations is verified by Monte Carlo simulations in Section \ref{Sec_Num_Res}. For brevity, we provide the SER results only for the M-PSK scheme.

\subsection{Asymptotic SNR analysis}
The approximation of the SER can be expressed in terms of diversity and coding gains by applying an asymptotic SNR analysis. In a high SNR region, the SER expression can be approximated as follows~\cite{giannakis}
\begin{equation}
\label{eqn:Asym_SER}
\bar{P}^\infty(\text{e})~\approx~\left ( G_{c}~\bar{\gamma} \right )^{-G_{d}},
\end{equation}
where \( G_d \) and \( G_c \) are the diversity and coding gains, respectively. 
\IEEEpubidadjcol

\begin{theorem}
\label{theo:approx_idn}
By utilizing the asymptotic property of the confluent hypergeometric Tricomi's function~\cite[eq.~(13.5.6)]{Abramowitz} for \( N_{\text{RIS}} > N_t + 2 \), the MGF of the instantaneous SNR \( \rho \) can be approximated as follows
\begin{equation}
    \label{asym_tricomi}
    \begin{aligned}
    M_{\rho}(-s) \approx \frac{\Gamma\left[N_{\text{RIS}}-N_t\right]}{\Gamma\left[N_{\text{RIS}}\right]}\left(\mu(s)\right)^{-N_t}\left(\beta(\phi)\right)^{-2N_t}.
    \end{aligned}
\end{equation}
By substituting the approximation in (\ref{asym_tricomi}) into (\ref{eqn:SER_PSK_Int}) and performing some mathematical simplifications, the asymptotic SER expression for M-PSK modulation can be expressed as

\begin{equation}
\begin{aligned}
\label{Asym_PSK}
    &\bar{P}_{\text{M-PSK}}^\infty(\text{e}) \approx \\ &\left[\frac{\beta^{2}(\phi)\alpha_{\text{PSK}}}{2 R_c}\bar{\gamma}\left(\frac{\Gamma(N_{\text{RIS}}-N_{t})}{\pi\Gamma(N_{\text{RIS}})}I(\theta)\right)^{-1/{N_t}}\right]^{-N_{t}},
\end{aligned}
\end{equation}
where
\begin{equation}
I\left(\theta\right) = \int_{0}^{\frac{(M-1)\pi}{M}}\sin^{2N_t}\left(\theta\right)~d\theta.
\end{equation}
Using (\ref{eqn:Asym_SER}) and (\ref{Asym_PSK}), the diversity and coding gain expressions can be provided as
\begin{subequations}
\begin{align}
G_{d} &= N_{t}, \label{eqn:div_order} \\
G_{c} &= \frac{\beta^{2} (\phi)\alpha_{\text{PSK}}}{2 R_c}\left(\frac{\Gamma(N_{\text{RIS}}-N_{t})}{\pi\Gamma(N_{\text{RIS}})}I\left(\theta\right)\right)^{-1/{N_t}} \label{eqn:code_gain}.
\end{align}    
\end{subequations}
\end{theorem}

It can be observed from (\ref{eqn:div_order}) and (\ref{eqn:code_gain}) that, under ideal conditions, the diversity order depends solely on the number of transmit antennas and is unaffected by other system parameters.  In contrast, the coding gain is significantly affected by the amplitude response coefficients of the RIS elements. Specifically, when the reflection amplitudes are relatively small or inconsistent among RIS elements, the overall coding gain deteriorates. This degradation becomes more apparent as the number of RIS elements increases, ultimately resulting in the degradation of the system error performance. Therefore, while RIS impairments do not affect spatial diversity, they critically influence the achievable coding gain and error performance.

\section{ERROR RATE ANALYSIS FOR NON-IDENTICAL RIS AMPLITUDE COEFFICIENTS}
\label{Sec_Per_An_Noniden}
In practical RIS deployments, the amplitude reflection coefficients of individual RIS elements are not necessarily identical due to hardware imperfections or design constraints.
This leads to independent and non-identical fading channel configurations, making the analysis considerably more challenging.
Although some statistical distributions exist in the literature to model such scenarios, they become mathematically intractable as the number of RIS elements \( N_{\text{RIS}} \) increases. In this study, we first show that the error rate performance can be exactly characterized using the hypoexponential distribution for small values of \( N_{\text{RIS}} \). However, as \( N_{\text{RIS}} \) grows, the complexity and limitations of the hypoexponential model make it impractical. To overcome this, we propose a novel approach based on the SPA technique to efficiently compute the error rates for arbitrary \( N_{\text{RIS}} \) under non-identical amplitude coefficients.
\IEEEpubidadjcol

\subsection{Error Rate Analysis in Low-\(N_{\mathrm{RIS}}\) Scenario}
\label{subsec:mgf_rho_nidn}

In scenarios where the amplitude coefficients of the RIS elements are not identical, the system exhibits non-identical and independent behaviors, which increases analytical complexity significantly. Specifically, the non-zero eigenvalue of the matrix \( \boldsymbol{\Phi}^{\dagger} \mathbf{g}^{\dagger} \mathbf{g} \boldsymbol{\Phi} \), denoted by \( \lambda \), becomes a sum of exponentially distributed random variables with different rate parameters \( \frac{1}{\beta^{2}_{i}(\phi_{i})} \). This sum does not follow an Erlang distribution as in the i.i.d. case, but rather a more general hypo-exponential distribution, as stated in the following theorem.

\begin{theorem}
\label{theo:f_lamb_nidn}
From (\ref{summ_exp}), \( \lambda = \sum_{i=1}^{N_{\text{RIS}}} X_i \), where \( X_i \sim \text{Exp}\left( \frac{1}{\beta^{2}_{i}(\phi_{i})} \right) \) are mutually independent exponential random variables with distinct rate parameters. Then, \( \lambda \) follows a hypo-exponential distribution~\cite{hypoexp}, and its PDF is given by
\begin{equation}
\label{lamb_hypo}
f_{\lambda}(y) = \left( \prod_{i=1}^{N} \frac{1}{\beta_i^2(\phi_i)} \right) 
\sum_{j=1}^{N} \frac{e^{-y / \beta_j^2(\phi_j)}}{\prod_{\substack{k=1 \\ k \neq j}}^{N} \left( \frac{1}{\beta_k^2(\phi_k)} - \frac{1}{\beta_j^2(\phi_j)} \right)}.
\end{equation}
\end{theorem}
\begin{corollary}
\label{coro:fz_nidn}
By substituting Equations~\eqref{lamb_hypo} and~\eqref{pdf_z_beta} into the expectation expression in~\eqref{eqn:fz_app}, and following similar mathematical simplifications, the exact PDF of \( Z \) can be expressed as
\begin{equation}
\begin{aligned}
\label{eqn:fz_nonid}
f_{Z}(a) =& \frac{2 \cdot a^{\frac{N_{t}-1}{2}}}{\Gamma(N_{t})} \cdot 
\left( \prod_{i=1}^{N_{\text{RIS}}} \frac{1}{\beta_{i}^{2}(\phi_{i})} \right) 
\\ \times& \sum_{j=1}^{N_{\text{RIS}}} \frac{\left( \beta_{j}^{2}(\phi_{j}) \right)^{\frac{1 - N_{t}}{2}}K_{1 - N_{t}}\left(2\sqrt{\frac{a}{\beta_{j}^{2}(\phi_{j})}}\right)}
{\prod_{\substack{k=1 \\ j \ne k}}^{N_{\text{RIS}}} \left( \frac{1}{\beta_{k}^{2}(\phi_{k})} - \frac{1}{\beta_{j}^{2}(\phi_{j})} \right) }.
\end{aligned}
\end{equation}
\end{corollary}
~

~

\begin{corollary}
\label{coro:mgf_rho_nidn}
Finally, by employing Equation~\eqref{eqn:fz_nonid} and applying the same MGF derivation procedure as in Section~\ref{subsec:mgf_rho}, the MGF of the instantaneous SNR \( \rho \) can be derived as
\begin{equation}
\label{eqn:mgf_rho_nidn}
\begin{aligned}
M_{\rho}(-s) &= \frac{1}{{\mu(s)}} 
\left( \prod_{i=1}^{N_{\text{RIS}}} \frac{1}{\beta_{i}^{2}(\phi_{i})} \right)
\\ &\times\sum_{j=1}^{N_{\text{RIS}}} 
\frac{U\left(1, 2 - N_t, \frac{1}{\mu(s) \beta_{j}^{2}(\phi_{j})} \right)}
{\prod_{\substack{k=1 \\ j \ne k}}^{N} 
\left( \frac{1}{\beta_{k}^{2}(\phi_{k})} - \frac{1}{\beta_{j}^{2}(\phi_{j})} \right)},
\end{aligned}
\end{equation}
\end{corollary}
Equation~\eqref{eqn:mgf_rho_nidn} provides an exact closed-form expression for the MGF of the instantaneous SNR under non-identical RIS amplitude coefficients. The SER expression for PSK modulation can be directly obtained by substituting this result into~\eqref{eqn:SER_PSK_Int}.
\IEEEpubidadjcol

\subsection{Error Rate Analysis in High-\(N_{\mathrm{RIS}}\) Scenario}
\label{sec:LCLT_and_SAP}

When the amplitude response coefficients of the RIS elements are non-identical, the PDF of \(\lambda\) follows a hypo-exponential distribution. As a result, the MGF of the instantaneous SNR also inherits a similarly complex structure, involving both hypo-exponential characteristics and the Tricomi's confluent hypergeometric function. This formulation results in significant mathematical complexity, making the SER analysis intractable, especially when the number of RIS elements is large. Therefore, in high \( N_{\text{RIS}} \) regimes, approximate analytical techniques must be employed to evaluate \( f_{\lambda}(y) \).  In this work, we consider two such approximation methods: SPA and LCLT. Both approaches are evaluated through mathematical derivations and validated through simulations. The results indicate that while both SPA and LCLT provide accurate approximations for large \( N_{\text{RIS}} \), the SPA method offers a significantly better approximation of \( f_{\lambda}(y) \), particularly for smaller RIS sizes. Therefore, using the SPA-based approximation of \( f_{\lambda}(y) \) is more suitable for deriving unified and accurate MGF and SER expressions.

\subsubsection{LCLT-based \( f_{\lambda}(y) \)}
\label{subsec:LCLT-f}

The LCLT method provides a tighter approximation when a large number of RIS elements, \(N_{\text{RIS}}\), is available.

\begin{theorem}
\label{theorem:approx_gauss_lambda}
Let \( \lambda = \sum_{i=1}^{N_{\text{RIS}}} X_i \), where \( X_i \sim \text{Exp}\left( \frac{1}{\beta_i^2(\phi_i)} \right) \) are mutually independent exponential random variables with finite means and variances. For large \( N_{\text{RIS}} \), the random variable \( \lambda \) converges to a Gaussian random variable
\begin{equation}
\lambda \overset{d}{\longrightarrow} \mathcal{N}\left( \mu_{\lambda}, \sigma_{\lambda}^2 \right),
\end{equation}
with
\begin{equation}
\mu_{\lambda} = \sum_{i=1}^{N_{\text{RIS}}} \beta_i^2(\phi_i)\quad\text{and}\quad \sigma_{\lambda}^2 = \sum_{i=1}^{N_{\text{RIS}}} \beta_i^4(\phi_i).
\end{equation}
Thus, the approximate PDF of \( \lambda \) can be expressed as

\begin{equation}
f_{\lambda}(y) \approx \frac{1}{\sqrt{2\pi \sigma_{\lambda}^2}} \exp\left( -\frac{(y - \mu_{\lambda})^2}{2\sigma_{\lambda}^2} \right).
\end{equation}
\end{theorem}

\noindent\textit{Note: The detailed conditions and derivation are provided in Appendix~\ref{appendix:proof_gaussian}}

\subsubsection{SPA-based \( f_{\lambda}(y) \)}
\label{subsec:SPA-f}
As an alternative to the LCLT, the SPA offers a highly accurate method for approximating the PDF of \( \lambda \). By utilizing the cumulant generating function (CGF) of the sum of independent random variables, SPA achieves tight approximations even when \( N_{\text{RIS}} \) is small.
\IEEEpubidadjcol

\begin{theorem}
\label{theorem:approx_saddle_lambda}
Let \( \lambda = \sum_{i=1}^{N_{\text{RIS}}} X_i \), where \( X_i \sim \text{Exp}\left( \frac{1}{\beta_i^2(\phi_i)} \right) \) are mutually independent exponential random variables with distinct parameters. Then, the PDF of \( \lambda \) can be approximated using the SPA as follows
\begin{equation}
\label{eqn:saddle_lambda}
  f_{\lambda}(y) \approx \frac{1}{\sqrt{2\pi \psi^{\prime\prime}(\hat{s})}} \exp\left( \psi(\hat{s}) - \hat{s}y \right),
\end{equation}
where \( \hat{s} \) is the saddle point corresponding to the solution of \( \psi'(\hat{s}) = y \), and \( \psi(s) \) is the CGF, defined as
\begin{equation}
\psi(s) = \log \mathbb{E}\left[ e^{s \lambda} \right] = -\sum_{i=1}^{N_{\text{RIS}}} \log(1 - s \beta^2_i(\phi_i)).
\end{equation}
The terms \( \psi(\hat{s}) \) and \( \psi^{\prime\prime}(\hat{s}) \) represent the CGF evaluated at the saddle point and its second derivative, respectively.
\end{theorem}
\noindent{\textit{Note: For a detailed derivation, refer to Appendix~\ref{appendix:proof_saddle}.}}
\vspace{6pt}

\begin{figure*}[htbp]
\centering
    \subfigure[$N_{\text{RIS}}=10$] {\includegraphics[width=2.25in]{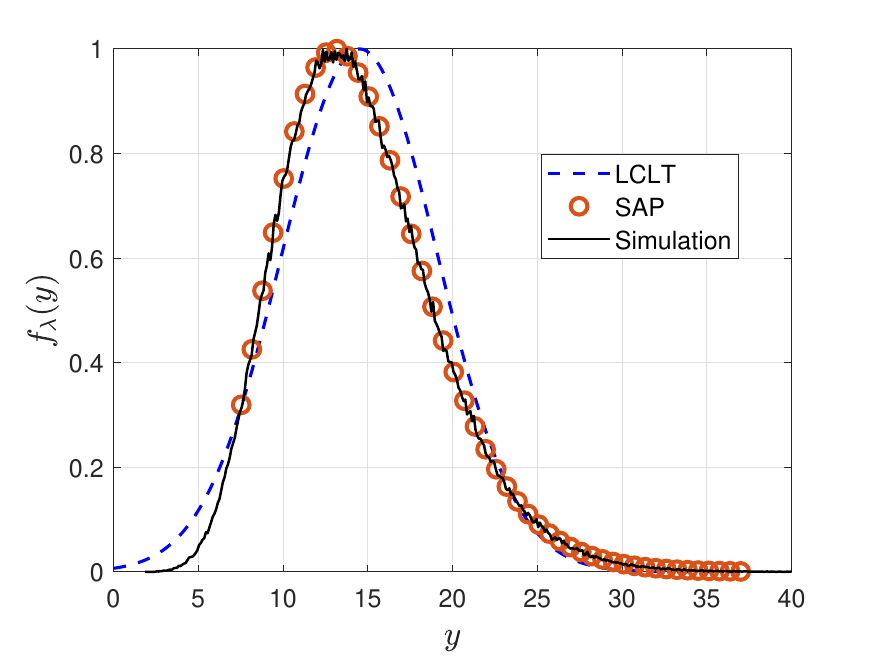}}
    \subfigure[$N_{\text{RIS}}=20$]{\includegraphics[width=2.25in]{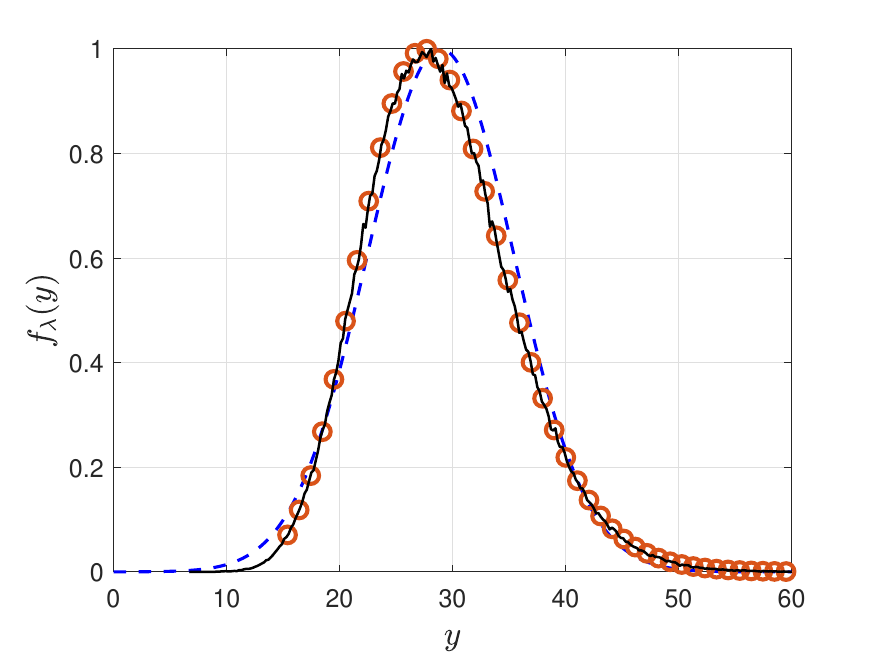}}
    \subfigure[$N_{\text{RIS}}=50$]
    {\includegraphics[width=2.25in]{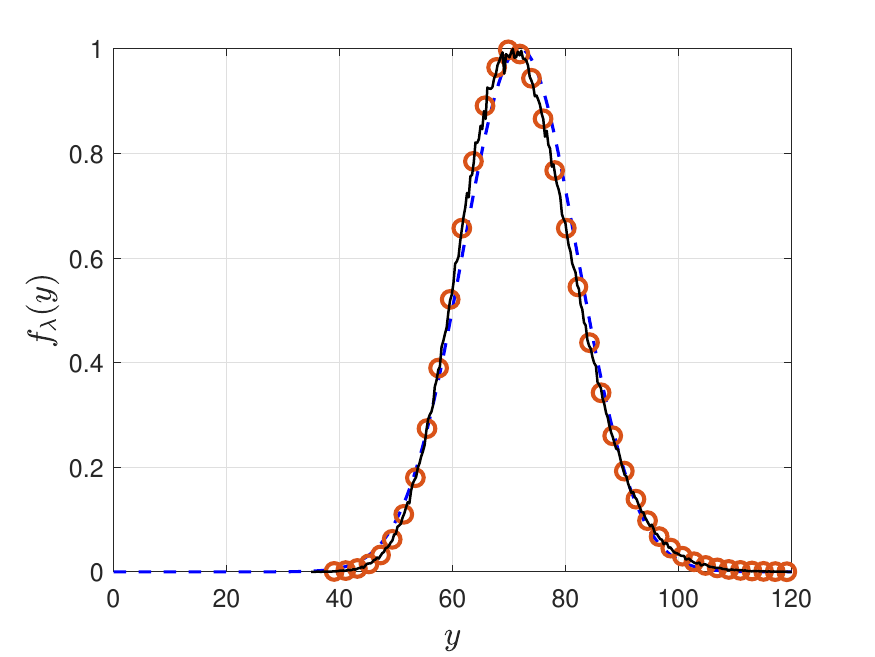}}
    \caption{Comparison of the LCLT and SPA-based approximation of \(f_{\lambda}(y)\) for different numbers of RIS elements, where \( \phi_n\in\left\{ 0, \frac{\pi}{2},\pi , \frac{3\pi}{2} \right\} \), \(\zeta_{\min}=0.8\), \(c = 0.43\pi\), and \(k=1.6\).}
\label{fig:Comp-Sad-LCLT}
\end{figure*}

SPA enables accurate estimation of the PDF for a wide range of \( N_{\text{RIS}} \), including scenarios where LCLT fails to provide sufficient precision. As illustrated in Fig.~\ref{fig:Comp-Sad-LCLT}, the SPA method exhibits a closer match to the empirical Monte Carlo results compared to the LCLT approximation, particularly in cases where \( N_{\text{RIS}} = 10 \) and \( N_{\text{RIS}} = 20 \). The difference between the LCLT approximation and the actual distribution is the most noticeable near the peak and in the tail regions. As $N_{\text{RIS}}$ increases, both approximations become more accurate: 
SPA remains very close to the simulated distribution, while LCLT gradually converges. This confirms the theoretical prediction that LCLT becomes increasingly accurate with larger \( N_{\text{RIS}} \), while SPA retains its accuracy even at moderate dimensions.

\subsubsection{Calculation of \(f_{Z}(a)\) and \(M_{\rho}(-s)\) with SPA-Based \(f_{\lambda}(y)\)}
\vspace{8pt}
As concluded in Sections~\ref{Sec_Per_An_Noniden}-\ref{sec:LCLT_and_SAP}\ref{subsec:LCLT-f} and~\ref{subsec:SPA-f}, the SPA-based approximation of \(f_{\lambda}(y)\) offers higher precision compared to the LCLT. Therefore, in the subsequent analysis, the SPA-based expression of \(f_{\lambda}(y)\) is used to derive \(f_{Z}(a)\) and\(M_{\rho}(-s)\). These quantities are essential for evaluating the error performance of RIS-assisted multiple-antenna systems with practical, non-identical amplitude response coefficients in the high-\(N_{\text{RIS}}\) regime. In addition, the SPA-based formulation yields consistently accurate results not only in the high-\(N_{\text{RIS}}\) regime but also for small and moderate values of \(N_{\text{RIS}}\).

\begin{corollary}
\label{coro:fz_nidn_sad}
By substituting Equations~\eqref{pdf_z_beta} and~\eqref{eqn:saddle_lambda} into the expectation expression in~\eqref{eqn:fz_app}, and following similar mathematical simplifications, the SPA-based \( f_Z(a) \) with numerical integral can be given as
\begin{equation}
    \label{eqn:f_z_saddle}
    f_{Z}(a) = \frac{a^{N_t-1}}{\sqrt{2\pi }\Gamma\left(N_{t}\right)}\int_{0}^{\infty}\frac{y^{-N_t}e^{\psi\left(\hat{s}\right)-\hat{s}y-\frac{a}{y}}}{\sqrt{\psi^{''}\left(\hat{s}\right)}}dy.
\end{equation}
\end{corollary}

\begin{corollary}
Finally, by employing Equation~\eqref{eqn:fz_nonid} and applying the same MGF derivation procedure as in Sections~\ref{subsec:mgf_rho} and~\ref{subsec:mgf_rho_nidn}, and benefiting from~\cite[eq.~(3.381.4)]{Ryzhik}, the SPA-based \( M_\rho(-s) \) can be derived as
\begin{equation}
\label{eqn:mgf_rho_nidn_sad}
\begin{aligned}
M_{\rho}(-s) = \frac{1}{{\sqrt{2\pi}}} \int_{0}^{\infty}\frac{e^{\psi\left(\hat{s}\right)-\hat{s}y}}{\sqrt{\psi^{''}\left(\hat{s}\right)}} 
\left(1+\mu(s)y\right)^{-N_t}dy.
\end{aligned}
\end{equation}
\end{corollary}

Due to the dependence of the saddlepoint \( \hat{s} \) on the integration variable \( y \) through the implicit equation \( \psi'(\hat{s}) = y \), the expression in~\eqref{eqn:mgf_rho_nidn_sad} cannot be expressed in an exact form. Consequently, the moment-generating function \( M_{\rho}(-s) \) must be evaluated numerically using appropriate quadrature techniques. The SER expressions for PSK modulation can be directly obtained by substituting this result into ~\eqref{eqn:SER_PSK_Int}. Importantly, this formulation enables the accurate computation of error rates for RIS-assisted systems with arbitrary numbers of reflecting elements, whether low or high. The proposed SPA-based solution yields results that closely match Monte Carlo simulations, effectively behaving as an exact solution. This observation is further confirmed and validated in Section \ref{Sec_Num_Res} through Monte Carlo simulations.
\subsubsection{Asymptotic SNR Analysis}
\label{sec:asym_nidn_neg_mom}
\begin{theorem}
\label{theorem:Asym_Neg_Mom}In the high-SNR regime \(\left( \text{i.e.},\mu(s)\rightarrow\infty\right)\), \( M_{\rho}(-s) \) can be given by the following asymptotic representation

\begin{equation}
\label{eqn:asym_rho_nidn}
\begin{aligned}
M_{\rho}(-s) = &\left(\frac{s\bar{\gamma}}{R_c}\right)^{-N_t}\frac{1}{\sqrt{2\pi}} \int_{0}^{\infty} \frac{y^{-N_t} e^{\psi(\hat{s}) - \hat{s} y}}{\sqrt{\psi''(\hat{s})}} dy
\\=&\left(\frac{s\bar{\gamma}}{R_c}\right)^{-N_t}\mathrm{E}_{\lambda}\left [y^{-N_t}  \right ].
\end{aligned}
\end{equation}
By substituting this expression into Equation~\eqref{eqn:SER_PSK_Int}, the asymptotic SER for M-PSK modulation under SPA-based modeling becomes
\begin{equation}
\label{SER_Asym_nidn}
\begin{aligned}
&\bar{P}_{\text{M-PSK}}^\infty(\text{e}) \approx \\&
\left[ \frac{\alpha_{\text{PSK}}}{2 R_c} \bar{\gamma} \left( \frac{I(\theta)}{\pi} 
\mathrm{E}_{\lambda}\left [y^{-N_t}  \right ]\right)^{-1/{N_t}} 
\right]^{-N_t}.
\end{aligned}
\end{equation}
Using Equation~\eqref{eqn:Asym_SER}, the diversity and coding gains can be identified from~\eqref{SER_Asym_nidn} as
\begin{subequations}
\begin{align}
    G_d &= N_t, \label{eqn:div_order_nidn} \\
    G_c &= \frac{\alpha_{\text{PSK}}}{2 R_c} \left( \frac{I(\theta)}{\pi} \mathbb{E}_{\lambda} \left[ y^{-N_t} \right] \right)^{-1/N_t} \label{eqn:code_gain_nidn} \\
        G_c &\propto \left( \mathbb{E}_{\lambda} \left[ y^{-N_t} \right] \right)^{-1/N_t}.
\end{align}
\end{subequations}
\end{theorem}

Interestingly, both Equation~\eqref{eqn:div_order} (for identical amplitude coefficients) and Equation~\eqref{eqn:div_order_nidn} (for non-identical amplitude coefficients) indicate that the diversity order \(G_d\) is solely determined by the number of transmit antennas \(N_t\), and remains unaffected by the RIS amplitude response. In contrast, the coding gain \(G_c\) is explicitly influenced by the distribution of \( \beta_i(\phi_i) \) through the cumulant generating function \( \psi(s) \), making it sensitive to the design and physical characteristics of the RIS. This observation confirms that, while RIS-induced amplitude fluctuations have no impact on spatial diversity in the asymptotic regime, they critically influence the effective coding gain. This behavior is confirmed through simulations presented in the subsequent section, where the theoretical asymptotic expressions exhibit a near-exact match with empirical SER curves.
\IEEEpubidadjcol

Building on this analytical insight, Equation~\eqref{eqn:code_gain_nidn} reveals a direct inverse relationship between the coding gain \(G_c\) and the \(N_t\)-th negative moment of the PDF of the random variable \( \lambda \). This connection serves as the key design principle for RIS phase shift optimization. Instead of directly minimizing the SER, an approach that typically requires complex analytical derivations or extensive simulations, we propose minimizing the negative moment \( \mathbb{E}_{\lambda}\left[ y^{-N_t} \right] \)~\cite{Neg_Moment, Mgf_Neg_Moment}, which serves as a mathematically grounded objective function for optimization. The proposed algorithm efficiently identifies RIS phase configurations that enhance coding gain and reduce SER with low complexity.

\begin{algorithm}[t]
\caption{Proposed Phase Shift Optimization Algorithm Using \(N_{t}\)-th Negative Moments of SPA-Based \(f_{\lambda}(y)\) with Group-Wise Greedy Search} \label{alg:greedy_groups}
\begin{algorithmic}[1]
\State \textbf{Input:} \( N_{\text{RIS}} \), \(N_t\), \(G\), \(T\), \(\Xi_b\), \( \mathbf{g} \)
\State Divide the RIS into $G$ disjoint groups: ${ \mathcal{G}_1, \mathcal{G}_2, \ldots, \mathcal{G}_G }$, where each group $\mathcal{G}_g$ contains $N_{\text{RIS}}/G$ consecutive elements
\State Initialize the random RIS phase configuration \(\mathbf{\Phi}\) by sampling each element from $\Xi_b$
\For{each group $g = 1, \ldots, G$}
\State Fix the phases for all elements not in group $\mathcal{G}_g$
\State Generate $T$ candidate phase vectors: \Statex \hspace{1.5em}${ \boldsymbol{\phi}_{g}^{(1)},\boldsymbol{\phi}_{g}^{(2)},\ldots,\boldsymbol{\phi}_{g}^{(T)}}$ for group $\mathcal{G}_g$, where each $\boldsymbol{\phi}_{g}^{(t)}$ \Statex \hspace{1.5em}is formed by randomly drawing values from $\Xi_b$
\For{each candidate index $t$ from 1 to $T$}
\State Construct a temporary RIS configuration $\mathbf{\Phi}_{t}$ by \Statex \hspace{3em}inserting $\boldsymbol{\phi}_{g}^{(t)}$ into the $\mathcal{G}_g$ portion of $\mathbf{\Phi}$
\State Compute reflection amplitudes for all elements \Statex \hspace{3em}using (\ref{Amp_Response})
\State Define RV \(\lambda_t\) by using~\eqref{summ_exp}
\State Compute the MGF \( M_{\lambda_t}(s) \) as defined in~\eqref{App_B2}
\State Compute the CGF \( \psi_t(s) \) using~\eqref{App_B3}
\State Calculate derivatives $\psi_{t}'(s)$ and $\psi_{t}''(s)$
\State Solve for the saddle point $\hat{s}$ such that $\psi_{t}'(\hat{s}) = y$
\State Compute the SPA-based PDF of \(\lambda_t\): $f_{\lambda_{t}}(y)$
\State Evaluate the \(N_t\)-th negative moment: \( \mathbb{E}_{\lambda_t}[y^{-N_t}] \)
\EndFor
\State Select the best-performing candidate $\mathbf{\Phi}_t$ with the \Statex \hspace{1.5em}minimum negative moment value and update $\mathbf{\Phi}$ \Statex \hspace{1.5em}accordingly in group $\mathcal{G}_g$
\EndFor
\State \textbf{Output:} Optimized RIS configuration $\mathbf{\Phi}$
\end{algorithmic}
\end{algorithm}

\section{OPTIMIZATION ALGORITHM DESIGN FOR RIS PHASE SHIFT CONFIGURATION} 
\label{sec_optimization}
In real-world implementations of RIS, each reflecting element is subject to hardware limitations that restrict it to a finite set of discrete phase shift values. For instance, under a 2-bit quantization scheme, each RIS element can only assume one of four predefined phase values, typically \( \phi \in \left\{ 0, \frac{\pi}{2}, \pi, \frac{3\pi}{2} \right\} \). This contrasts with idealized RIS models that assume continuous phase control over \( [0, 2\pi) \), and significantly affects beamforming precision and system-level performance. However, it is more practical to assume discrete phase shifts for RIS units due to hardware constraints, which makes the continuous phase-shift assumption unrealistic; this limitation arises from actual hardware structures \cite{Active_RISs, RIS_Phs_Res}.

As discussed in earlier sections, practical RIS hardware often exhibits an amplitude-phase coupling effect, whereby the amplitude response \( \beta_n(\phi_n) \) of each element depends on the applied phase shift \( \phi_n \). Although configuring each RIS element with the phase value that maximizes its individual amplitude response (e.g., \( \phi_n = \pi/2 + c \)) might intuitively be expected to yield the lowest SER or the highest ergodic capacity, such a configuration is not practically feasible. This is because the phase shift that results in the maximum amplitude response may not be realizable due to the hardware limitations of practical RIS implementations. Consequently, in real systems, the phase profile of RIS elements must be carefully adjusted not only to maximize local amplitude responses but also to align with the overall objectives of the system.

To this end, we propose a low-complexity, yet effective, optimization framework that accounts for these hardware-induced constraints. Our approach exploits the insight that the coding gain \(G_c\) of the space-time coded RIS-assisted multiple antenna systems is strongly related to the \( N_t \)-th negative moment of the SPA-based \(f_{\lambda}(y)\). This observation enables us to formulate a mathematically grounded optimization criterion: minimizing the negative moment of the eigenvalue distribution corresponds to maximizing the coding gain, and thus improving SER performance.
\begin{figure}[t!]
    \centering {\includegraphics[width=3.8in, angle=0]{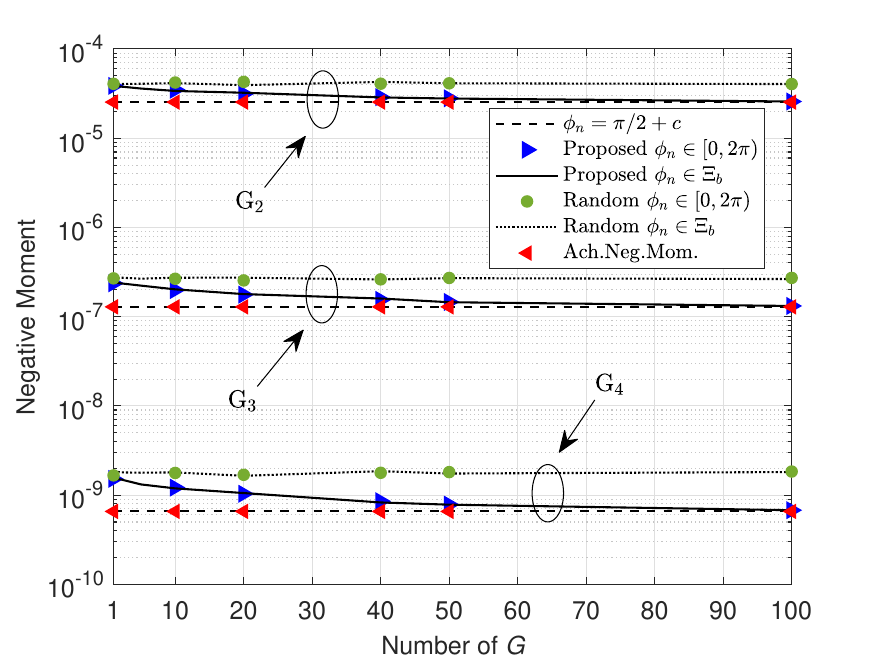}}
    \caption{Comparison of negative moment values achieved by the proposed algorithm, optimum phase configuration, and random phase assignment for different grouping levels \( G \) and \( N_{\mathrm{RIS}} = 200 \).}
    \label{fig.neg_mom}
\end{figure}

From~\eqref{eqn:code_gain}, we consider an ideal scenario in which all the RIS elements are perfectly configured with \( \beta^2(\phi) = 1 \). By taking the ratio between this ideal-case coding gain and that obtained under non-identical amplitude responses (e.g.,~\eqref{eqn:code_gain_nidn}), the relative coding gain degradation can be quantified as follows
\begin{equation}
\label{opt_rate}
r = \frac{\Gamma\left(N_{\mathrm{RIS}} - N_t\right)}{\Gamma\left(N_{\mathrm{RIS}}\right)\, \mathbb{E}_{\lambda}\left[y^{-N_t}\right]}.
\end{equation}

As can be seen from this expression, when the \(N_t\)-th negative moment \( \mathbb{E}_{\lambda}\left[y^{-N_t}\right] \) becomes equal or very close to \( \Gamma\left(N_{\mathrm{RIS}} - N_t\right)/\Gamma\left(N_{\mathrm{RIS}}\right) \), the resulting SER performance of the optimized configuration approaches that of the ideal phase configuration. Importantly, this ratio also reveals the theoretical lower bound of the negative moment expression. Therefore, the optimization algorithm effectively aims to minimize the negative moment until it asymptotically reaches its minimum possible value, which corresponds to the ideal SER performance. This mathematical insight provides a strong justification for the negative-moment-based criterion used in our optimization algorithm and further highlights its capability to achieve near-optimal performance under practical hardware constraints. Notably, this formulation eliminates the need for computationally intensive Monte Carlo-based SER evaluations during the optimization process. Furthermore, this approach enhances the practical deployability of the proposed method. In hardware-constrained systems, direct SER computation is often infeasible in real time. However, since the eigenvalue distribution, or its approximated PDF, can be inferred from observable signal power measurements, the negative moment serves as a meaningful and tractable surrogate metric for SER minimization in practical systems.
\begin{figure}[t!]
    \centering {\includegraphics[width=3.8in, angle=0]{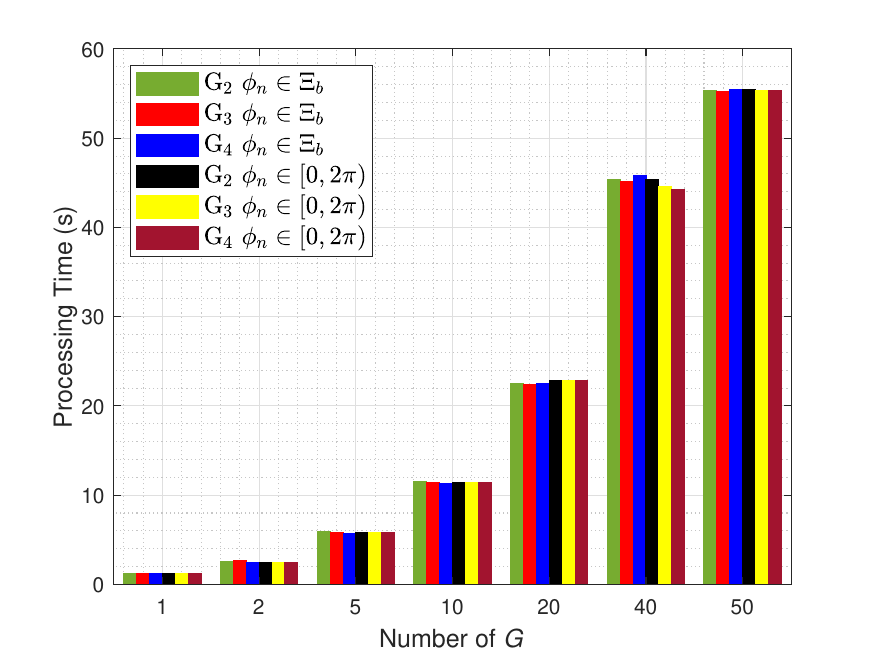}}
    \caption{Processing time required to determine the optimal phase configuration minimizing the \( N_t \)-th negative moment of the SPA-based \(f_{\lambda}(y)\) for \( N_{\mathrm{RIS}} = 200 \).}
    \label{fig.proc_time}
\end{figure}

Optimization is performed using a greedy group search strategy as provided in Algorithm~\ref{alg:greedy_groups}. The RIS is partitioned into \( G \) disjoint groups of consecutive elements, and the algorithm optimizes each group sequentially while holding the others fixed. For each group, a set of \( T \) candidate phase configurations is generated from the quantized codebook \( \Xi_b \). The quantized phase shift codebook is defined as \( \Xi_b = \left\{ 0, \frac{2\pi}{2^b}, \dots, \frac{2\pi(2^b -1)}{2^b} \right\} \), where \( b \) denotes the quantization resolution (in bits) of each RIS element. Each candidate configuration is evaluated by computing the phase-dependent amplitude responses and estimating the corresponding SPA-based \( f_\lambda(y) \).

\subsection{Performance Evaluation of the Proposed Algorithm}
\label{sec:opt_perf_eva}

Fig.~\ref{fig.neg_mom} illustrates the computed \( N_t \)-th negative moment values under various phase configuration schemes and grouping levels \( G \). The performance of the proposed optimization algorithm is compared against random phase assignment, with system parameters set to \(\zeta_{\min} = 0.8\), \( c = 0.43\pi \), \( k = 1.6 \), and assuming \( T = 10^6 \) candidate configurations. It can be observed that \( G_2 \) scheme achieves the achievable negative moment with fewer groups under both continuous and quantized phase configurations. This is particularly significant for practical systems, where only quantized phase shifts are feasible due to hardware constraints. This behavior suggests that for practical applications prioritizing higher data rates and where ultra-high reliability is not critical, the use of the proposed algorithm with the \( G_2 \) coding structure provides an effective balance between performance and computational efficiency. Moreover, as depicted in Figure~\ref{fig.proc_time}, this setting requires significantly less processing time. Conversely, in reliability-critical use cases such as industrial automation or mission-critical wireless systems, coding schemes like \( G_4 \) (i.e., with more transmit antennas and a lower code rate) become more desirable.

The performance gap between random phase assignment and the optimum configuration becomes significantly larger when both the number of RIS elements and the number of transmit antennas increase. This is because the achievable minimum negative moment value decreases substantially in such scenarios, as implied by the ratio expression in \eqref{opt_rate}.  Consequently, the optimization algorithm proves especially valuable in coding schemes such as \( G_4 \), which offer full diversity gain with low code rates, as achieving near-optimal performance becomes critical for high-reliability applications. In these settings, although the optimization process becomes slower due to the higher group count required to achieve the optimum performance, the algorithm’s capability to approach the theoretical performance bound—clearly illustrated in Fig.~\ref{fig.neg_mom}. This trade-off between performance and processing time is further quantified in Fig.~\ref{fig.proc_time}, highlighting the importance of selecting appropriate grouping levels in practice.
\begin{figure}[t!]
    \centering {\includegraphics[width=3.8in, angle=0]{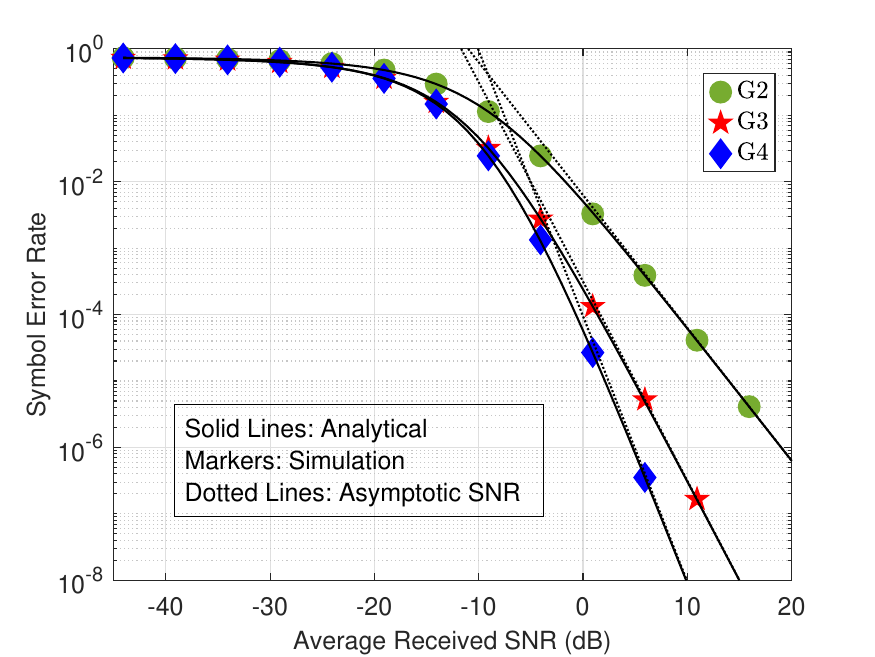}}
    \caption{SER performance for different OSTBC schemes, evaluated for \(N_{\mathrm{RIS}} = 32,~\phi_n = \pi\).}
    \label{fig.3}
\end{figure}

\textit{Note:} The primary goal of this study is to introduce and validate a theoretically grounded optimization objective. For proof-of-concept purposes, a basic Greedy Search algorithm is employed. In future work, more efficient AI-based optimization strategies can be developed using this objective function to enable real-time adaptation in practical RIS-assisted systems.

\section{NUMERICAL RESULTS AND INSIGHTS}
\label{Sec_Num_Res}
In this section, a realistic simulation environment is developed in \textsc{MATLAB} to validate the derived mathematical expressions and to demonstrate the impact of the proposed optimization algorithm on system performance. 
In addition, to illustrate the effects of RIS imperfections and optimization strategies on system error performance, numerical results are provided for various deployment scenarios.
\subsection{Path Loss Modeling and Simulation Parameters}
\IEEEpubidadjcol
The main focus of this study is not on comparing different path loss models but rather on providing a practical and realistic path loss formulation for RIS-assisted wireless systems. Given that, in the considered deployment scenarios, both the transmitter and the receiver are located in close proximity to the RIS, the near-field PL (NFPL) condition is predominantly satisfied. To justify this assumption, we observe that the distances between the RIS and the transceiver nodes are smaller than or comparable to the Fraunhofer distance, defined as
\begin{equation}
    d_\text{NF} = \frac{\text{max}(d_{M}, d_{N})}{l},
\end{equation}
where \(\text{max}(d_{M}, d_{N})\) denotes the largest physical dimension of the RIS and $l$ is the wavelength. For instance, with $\text{max}(d_{M}, d_{N}) \approx 4.0397~\text{m}$ (the maximum RIS length used in our setup) and $l = 0.0789\,\text{m}$ (corresponding to carrier frequency \(f_c = 3.8\,\mathrm{GHz}\)), the Fraunhofer distance is $d_\text{NF} = 413.6632\,\text{m}$.
\begin{figure}[t]
    \centering {\includegraphics[width=3.8in, angle=0]{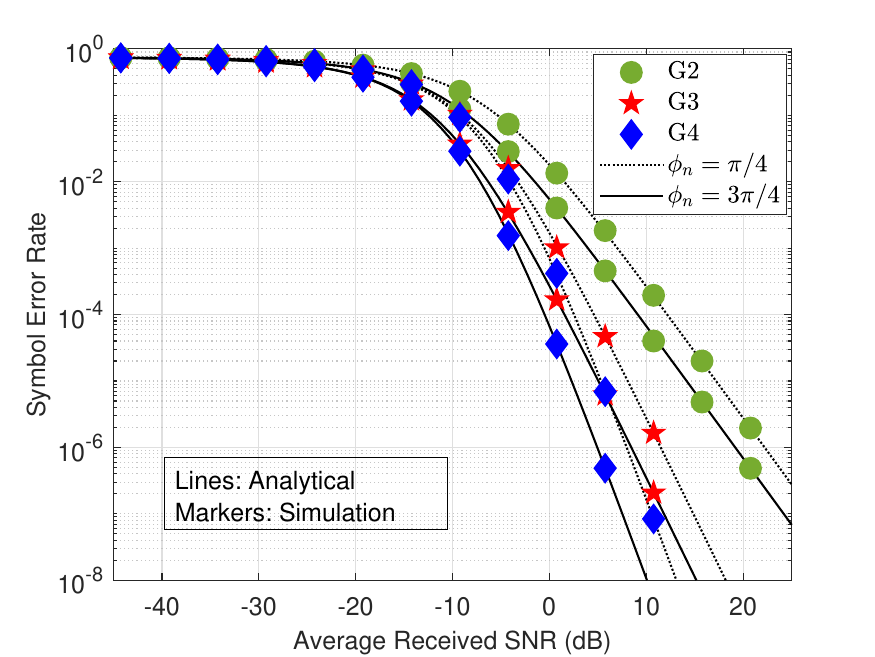}}
    \caption{SER performance for different OSTBC schemes and phase shift values (\( N_{\mathrm{RIS}} = 32 \)).}
    \label{fig.4}
\end{figure}

In the NFPL case, the cascaded channel between the transmitter and receiver via the RIS cannot be accurately modeled by the product of two independent links as in the far-field PL (FFPL) scenario.
According to the ETSI technical report~\cite{etsigr2023003111}, the total NFPL can be approximated as
\begin{equation}
\label{eq:PL_near}
P_{L} \approx G_t G_r \frac{l^2}{16\pi^2} \left[ \frac{\beta_{\mathrm{max}}}{(d_t + d_r)} \right]^2,
\end{equation}
where \( G_t \) and \( G_r \) denote the transmit and receive antenna gains, \( l \) is the wavelength, \( \beta_{\mathrm{max}} \in [0,1] \) represents the maximum amplitude reflection coefficient among RIS elements, and \( d_t \) and \( d_r \) are the distances from the transmitter and receiver to the RIS center, respectively. The distances $d_t$ and $d_r$ are computed from the deployment geometry as
\begin{equation}
d_t = \sqrt{d_{tx}^2 + d_{ty}^2}, \qquad
d_r = \sqrt{d_{rx}^2 + d_{ry}^2}.
\end{equation}
In the considered simulation scenarios, the coordinate differences are set as $d_{tx} = d_{rx} = 30~\mathrm{m}$ and $d_{ty} = d_{ry} = 40~\mathrm{m}$. Accordingly,
\(d_t = d_r = 50~\mathrm{m}\). The geometric meaning of these coordinate differences, such as $d_{tx}$ and $d_{ty}$, is clearly illustrated in the system model depicted in Fig.~1.
All subsequent theoretical derivations and simulation results in this work are based exclusively on the NFPL formulation given in \eqref{eq:PL_near}.
\IEEEpubidadjcol

\begin{figure}[t]
    \centering {\includegraphics[width=3.8in, angle=0]{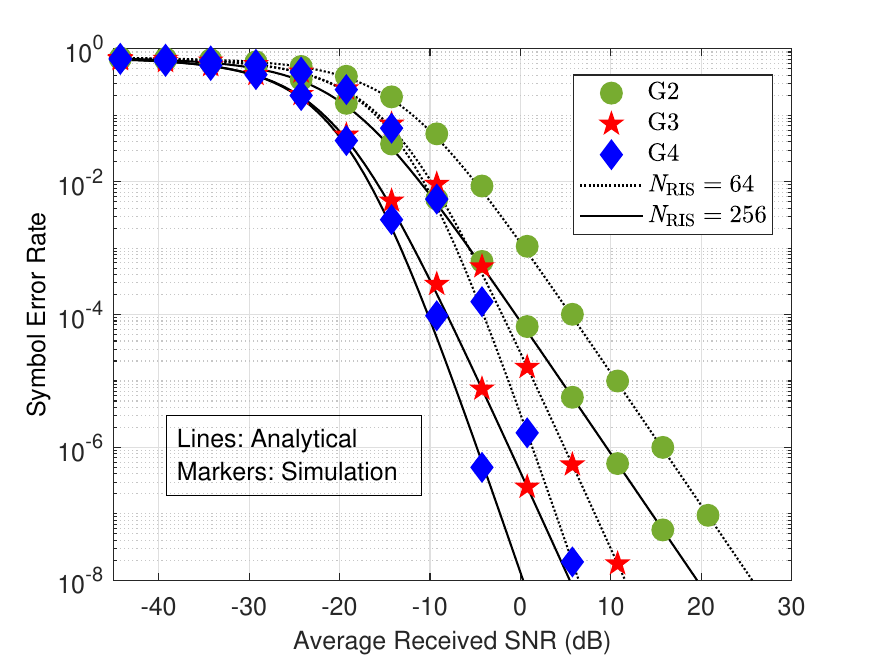}}
    \caption{SER performance for different \( N_{\mathrm{RIS}} \) values (\( \phi_n = 3\pi/4 \)).}
    \label{fig.5}
\end{figure}

For the simulations, 
the centers of the transmitter, receiver, and RIS are assumed to be aligned, i.e., $\varepsilon_t = \varepsilon_r = 0$, and the azimuth angles of the transmitter and receiver are set to $30^\circ$ and $120^\circ$, respectively. The other simulation parameters are as follows: $d_x = l/5$, $d_y = l/5$, $G_t = G_r = 1$. To ensure that both the simulations and analytical evaluations are consistent with practical RIS hardware constraints, we assume a 2-bit quantization model for each RIS element. Accordingly, the phase shift values are restricted to a discrete set, i.e., \( \phi_n \in \Xi_2\in \left\{ 0, {\pi/2}, {\pi}, 3\pi/2 \right\} \), except for Fig.~\ref{fig.4} and Fig.~\ref{fig.5}. It should be noted that this discrete set is applied solely in the quantized phase scenarios. Other cases in the paper, such as uniformly distributed phases, are treated separately in both analytical and simulation frameworks. Furthermore, in all simulations and analytical computations, the highest amplitude response and circuit parameters are set as follows: \(\zeta_{\min}=0.8\), \(c = 0.43\pi\), and \(k=1.6\).

\subsection{Performance Results for Identical Amplitude Response Case}
When all reflective elements have identical amplitude coefficients, the error performance calculations are presented in Section~\ref{Sec_Per_An_Iden}.  Fig.~\ref{fig.3} illustrates the SER performance of an RIS-assisted multiple-antenna system employing different OSTBC schemes, under the assumption that all RIS elements have identical reflection amplitude responses and a common phase shift value of \( \phi = \pi \). The figure presents both analytical results and Monte Carlo simulations, along with asymptotic SER curves to provide insights into the achievable diversity and coding gains. It is observed that the diversity gain is directly proportional to the number of transmit antennas and that full spatial diversity is achieved regardless of the specific OSTBC scheme employed. These results confirm the robustness of the system in preserving full diversity order, even in the presence of realistic RIS hardware constraints.
\IEEEpubidadjcol

In Fig.~\ref{fig.4}, the SER performance of the RIS-assisted system is presented for different OSTBC techniques under two distinct phase shift configurations. Unlike the scenario considered in the previous figure, here the impact of two different fixed phase values (\( \phi = \pi/4 \) and \( \phi = 3\pi/4 \)) applied uniformly to all RIS elements is investigated. The diversity order remains unaffected, and each coding scheme continues to achieve full spatial diversity. This observation is consistent with the theoretical expression given in \(\textbf{Theorem \ref{theo:approx_idn}} \), where the square of the phase shift coefficient influences the coding gain but does not affect the diversity order. The near-perfect agreement between the simulation and the theoretical results further validates the proposed analytical framework in the case of identical amplitude responses.

 \begin{figure}[t]
    \centering {\includegraphics[width=3.8in, angle=0]{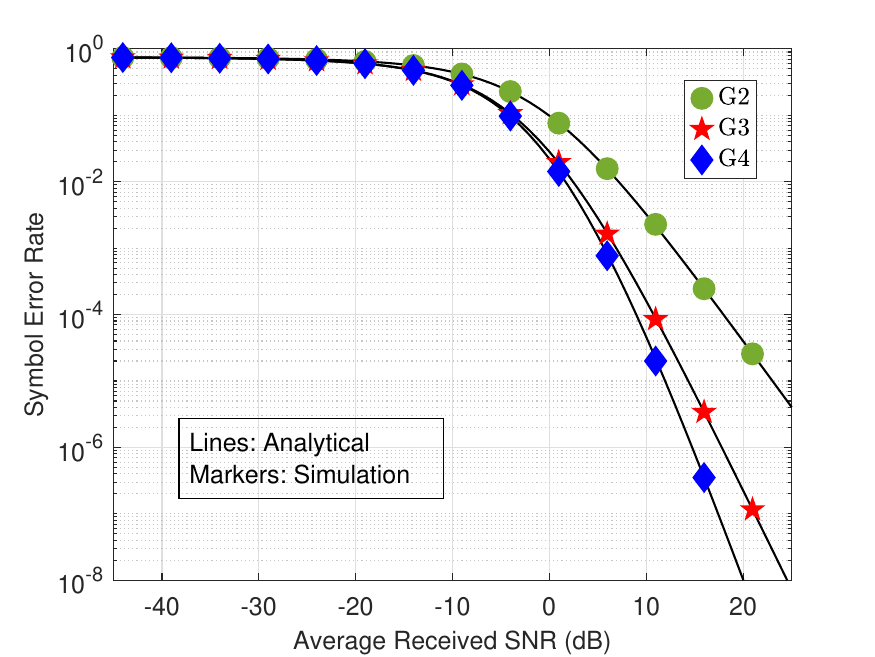}}
    \caption{SER performance for uniformly distributed phase shift values \( \phi_n \in [0, 2\pi) \), evaluated for \( N_{\mathrm{RIS}} = 6 \).}
    \label{fig.6}
\end{figure}

Fig.~\ref{fig.3} and Fig.~\ref{fig.4} illustrate the error performance for different reflection coefficient values, assuming a fixed number of RIS elements. These results are based on the closed-form SER expression, derived by substituting the exact MGF for the identical case~(\ref{mgf_z_Tricomi}) into the general SER formula~(\ref{eqn:SER_PSK_Int}). However, as discussed in previous sections, the presence of Tricomi's confluent hypergeometric function in the MGF introduces significant computational complexity. This makes the SER evaluation infeasible for large RIS sizes.

To overcome this limitation, the SPA-based MGF expression~(\ref{eqn:mgf_rho_nidn}) is used for large-scale RIS scenarios. This approximate MGF is then inserted into~(\ref{eqn:SER_PSK_Int}) to compute the SER. As shown in Fig.~\ref{fig.5}, the SPA-based SER results closely match the Monte Carlo simulations for various RIS sizes. Moreover, increasing the number of RIS elements improves the coding gain (evident from the 6~dB SNR gain when the RIS size is quadrupled), while the diversity gain remains solely dependent on the number of transmit antennas. The SPA-based method thus offers an accurate and computationally efficient tool for analyzing both identical and non-identical amplitude response RIS configurations.

\subsection{Performance Results for Non-identical Amplitude Response Case}
\label{Sec.Num_Res_Nonidn}
\IEEEpubidadjcol

In this subsection, comprehensive performance evaluations are provided for the non-identical amplitude response case, where the reflection amplitudes of RIS elements vary due to practical hardware constraints. Additionally, the effectiveness of the optimization algorithm proposed in the previous section is examined by comparing its SER performance against both the ideal optimum phase configuration and a randomly quantized phase configuration. The evaluation includes performance analysis with different group counts, RIS sizes, and OSTBC coding schemes. During the optimization process, the number of candidate phase configurations considered for each group is fixed to \( T = 10^{6} \), ensuring a sufficiently rich search space.
\begin{figure}[t!]
    \centering {\includegraphics[width=3.8in, angle=0]{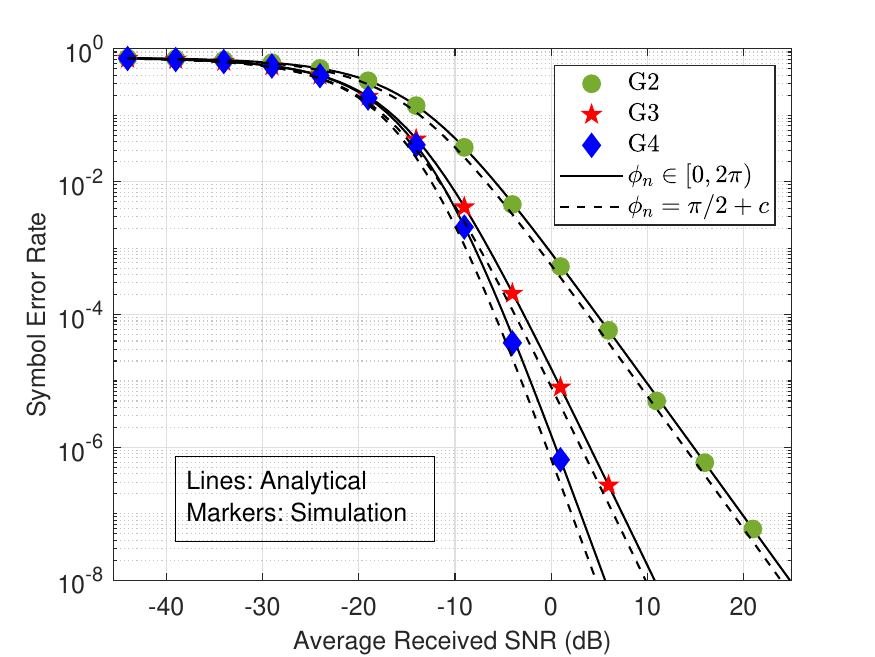}}
    \caption{SER performance for uniformly distributed and optimal phase shift values, evaluated for \( N_{\mathrm{RIS}} = 100 \).}
    \label{fig.7}
\end{figure}

Fig.~\ref{fig.6} illustrates the SER performance of the system for the case, where each RIS element has a distinct amplitude response. In this scenario, the phase shifts \( \phi_n \) are uniformly distributed over the interval \( [0, 2\pi) \), and the number of RIS elements is set to a small value of \( N_{\text{RIS}} = 6 \). The analytical curves are obtained using the exact MGF expression derived in \eqref{eqn:mgf_rho_nidn}, which models the distribution of the non-zero eigenvalue using hypoexponential structures and Tricomi’s confluent hypergeometric functions. Although the MGF expression is theoretically exact, its complexity, caused by the involvement of special functions, limits its applicability to relatively small RIS sizes. As shown in the figure, the analytical results closely match the Monte Carlo simulations, confirming the accuracy of the analysis. Additionally, it is observed that the diversity orders of different coding schemes remain the same as in the case of identical amplitude responses.
\IEEEpubidadjcol

\begin{figure}[t!]
    \centering {\includegraphics[width=3.8in, angle=0]{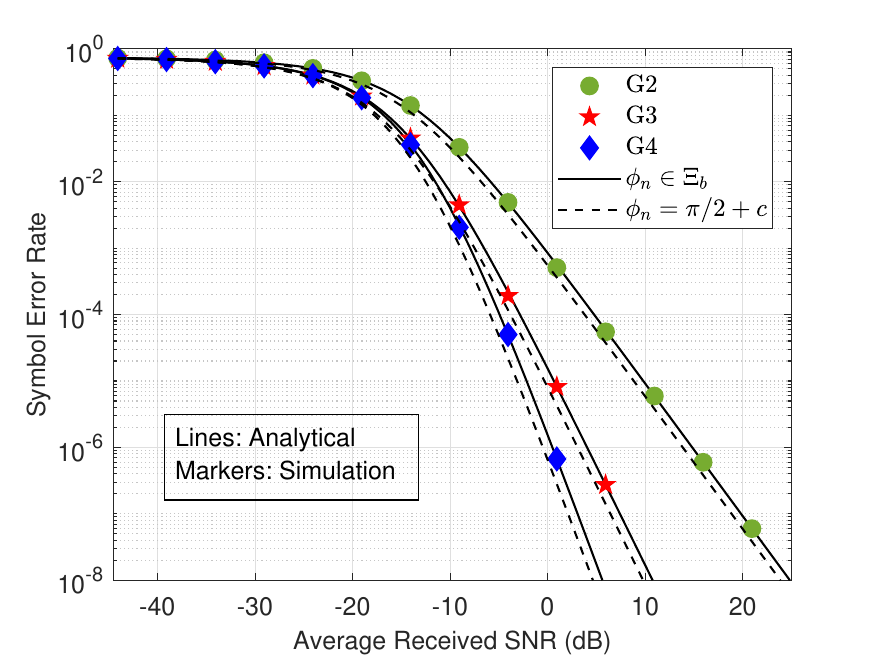}}
    \caption{SER performance for quantized and optimal phase shift values, evaluated for \( N_{\mathrm{RIS}} = 100 \).}
    \label{fig.8}
\end{figure}

\begin{figure}[t!]
    \centering {\includegraphics[width=3.8in, angle=0]{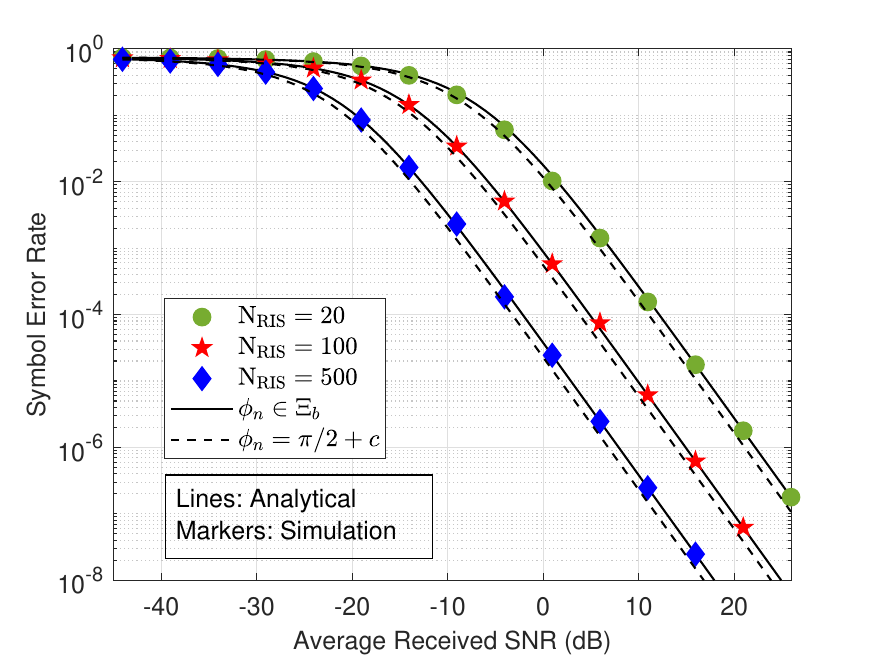}}
    \caption{SER performance RIS-assisted Alamouti STBC scheme under uniformly distributed and optimal phase shifts for different \( N_{\mathrm{RIS}}\).}
    \label{fig.9}
\end{figure}

Figs.~\ref{fig.7}--\ref{fig.9} illustrate the SER performance of the system under the non-identical amplitude response model. In Fig.~\ref{fig.7}, analytical and simulation-based SER curves are presented for large \( N_{\mathrm{RIS}} \) values, comparing two cases: (i) uniformly distributed phase shifts over \( [0, 2\pi) \), and (ii) fixed ``optimum'' phase shifts that maximize the amplitude response of each RIS element (i.e., where each RIS element \(\beta_n(\phi_n)\) is set to 1 for \(\phi_n=\pi/2+c\)). These results, obtained using the SPA-based approximation in~\eqref{eqn:mgf_rho_nidn_sad}, confirm that the proposed analytical model remains accurate for large-scale systems and provides a close match with Monte Carlo simulations.

Fig.~\ref{fig.8} extends this evaluation to practical RIS configurations, where each phase shift is randomly selected from a quantized codebook \( \Xi_b \). Despite the coarse phase quantization, the SER performance remains nearly identical to that of the ideal continuous-phase scenario, validating the effectiveness of the analytical expression and the robustness of the system under hardware-imposed constraints. The results also indicate that diversity and coding gains remain consistent with the identical amplitude response case.
\IEEEpubidadjcol

Fig.~\ref{fig.9} presents a comparative analysis for different values of \( N_{\mathrm{RIS}} \), focusing on the \( \mathbf{G}_2 \) coding scheme. It is observed that increasing the number of RIS elements significantly enhances the coding gain. Interestingly, for any given \( N_{\mathrm{RIS}} \), the performance gap between the optimum and quantized random phase configurations remains nearly constant at approximately 1~dB. However, this gap tends to increase in higher-order OSTBC schemes such as \( \mathbf{G}_3 \) and \( \mathbf{G}_4 \), as will be demonstrated in the subsequent results. This observation indicates that although the use of quantized phase profiles without optimization introduces a non-negligible degradation in SER performance, the severity of this degradation does not scale with the RIS size when the coding scheme is fixed. In other words, the impact of phase quantization remains effectively bounded for different RIS dimensions. Nevertheless, the existence of such a gap suggests that notable performance gains can still be achieved through effective RIS optimization.
\IEEEpubidadjcol

\begin{figure}[t!]
    \centering {\includegraphics[width=3.8in, angle=0]{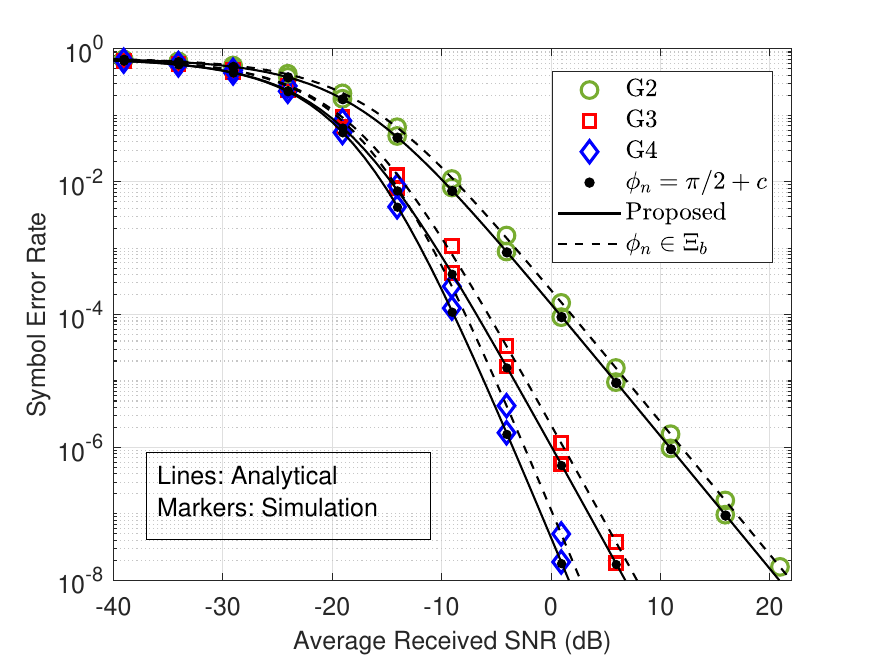}}
    \caption{SER performance comparison of the proposed optimization algorithm, random phase configuration, and optimum phase configuration for \( N_{\mathrm{RIS}} = 200 \) and \( G = 100 \).}
    \label{fig.10}
\end{figure}

Fig.~\ref{fig.10} presents the SER performance of the proposed negative \( N_t \)-th moment-based optimization algorithm, in comparison with both the optimum phase configuration and the quantized random phase configuration. The results are shown for a representative large-scale RIS scenario with 200 elements, partitioned into \( G = 100 \) groups.
As clearly illustrated, the proposed algorithm successfully achieves SER performance that closely matches the optimum configuration for each OSTBC technique, thereby validating its effectiveness in approximating the ideal RIS phase profile even under practical quantization constraints. Moreover, the simulation results match the analytical curves, confirming the accuracy and reliability of the proposed design.

\begin{figure}[t!]
    \centering {\includegraphics[width=3.8in, angle=0]{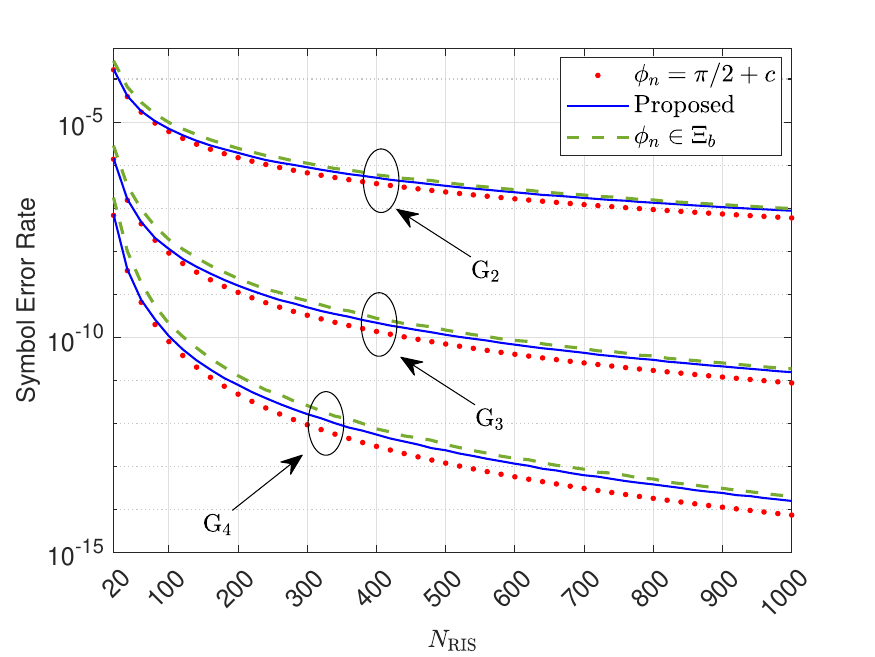}}
    \caption{SER performance comparison of the proposed optimization algorithm, random phase configuration, and optimum phase configuration for different values of \( N_{\mathrm{RIS}} \) for \( \bar{\gamma} \approx 10~\text{dB} \) and \( G = 20 \).}
    \label{fig.11}
\end{figure}

Fig.~\ref{fig.11} illustrates the SER performance as a function of \( N_{\mathrm{RIS}} \), for a fixed average received SNR and a fixed group count of \( G = 20 \). As seen in the figure, for relatively small values of \( N_{\mathrm{RIS}} \), the proposed optimization algorithm successfully matches the performance of the optimum phase configuration, despite the limited number of groups. \IEEEpubidadjcol However, as the number of RIS elements increases, the performance of the proposed algorithm gradually converges to that of the random quantized phase configuration. This is due to the fact that with a fixed and low group number, the resolution of phase optimization becomes insufficient to track the optimal configuration in large-scale RIS deployments. It should be noted that this performance gap becomes larger for higher-order OSTBCs, such as \( \mathbf{G}_4 \), compared to lower-order codes. This behavior can be theoretically explained in Section~\ref{sec_optimization}. As the number of transmit antennas \( N_t \) and RIS elements \( N_{\mathrm{RIS}} \) increase, the minimum achievable negative moment value decreases significantly. Consequently, under low group resolution, the configuration identified by the proposed algorithm yields a negative moment that remains considerably higher than the theoretical threshold, especially for large \( N_t \), resulting in a more noticeable SER gap from the optimum configuration. Therefore, the figure highlights a key trade-off between computational complexity and performance, emphasizing the importance of properly selecting the group count \( G \) when scaling the RIS size.

\begin{figure}[t]
    \centering {\includegraphics[width=3.8in, angle=0]{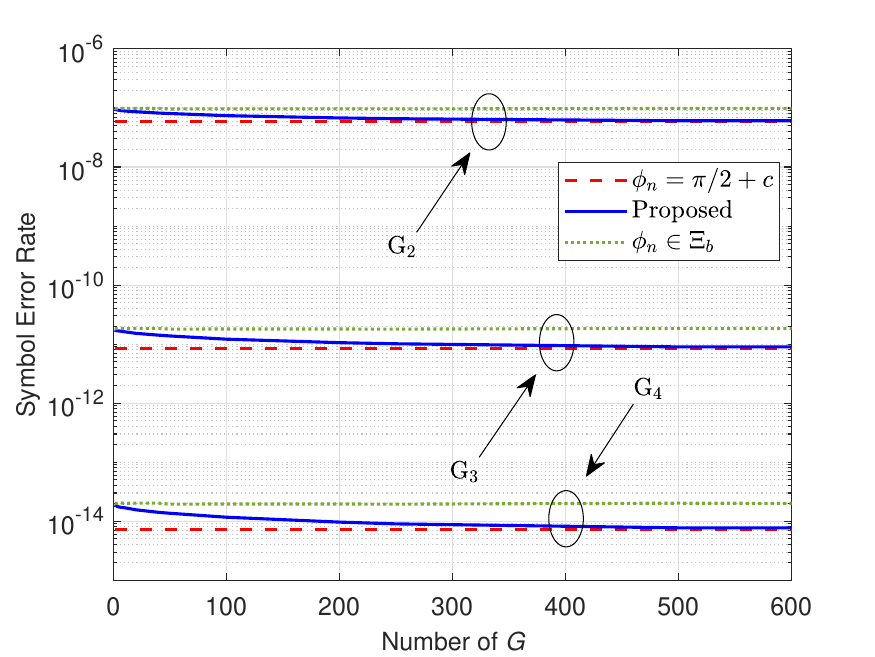}}
    \caption{SER performance comparison of the proposed optimization algorithm, random phase configuration, and optimum phase configuration for different values of \( G \), evaluated with \( \bar{\gamma} \approx 10~\text{dB} \) and \( N_{\mathrm{RIS}}=1000. \)}
    \label{fig.12}
\end{figure}

Fig.~\ref{fig.12} presents the SER performance of the proposed optimization algorithm as a function of the number of groups \( G \), for \( N_{\mathrm{RIS}} = 1000 \) and \( \bar{\gamma} \approx 10~\mathrm{dB} \).
As expected, for small values of \( G \) (i.e., large group sizes), the performance of the proposed algorithm closely resembles that of the random phase configuration due to limited phase resolution. As the group count increases, the proposed algorithm exhibits a rapid performance improvement and eventually converges to the optimum phase configuration. Moreover, consistent with previous findings, the performance gap between the proposed algorithm and the random configuration is more noticeable for higher-order OSTBCs. Consequently, the results in Fig.~\ref{fig.12} reaffirm that the proposed algorithm is particularly beneficial in such scenarios and emphasize the critical role of group resolution in achieving near-optimal performance in large-scale RIS-assisted systems.
\IEEEpubidadjcol

\section{CONCLUSION AND DISCUSSIONS}
\label{Sec_Conclusion}
This work presents a comprehensive SER analysis for space-time-coded RIS-assisted multiple-antenna systems under practical reflection models that include phase-dependent and nonidentical amplitude responses. Using the Gramian structure of the cascaded fading channel, we derive exact expressions for the PDF and MGF of the instantaneous SNR for systems with a small number of RIS elements. For large-scale RIS deployments, the resulting analytical complexity renders exact derivations intractable. To overcome this, we employ the SPA to obtain highly accurate approximations of the PDF and MGF, enabling a unified and tractable SER analysis for arbitrary RIS configurations. The simulation results demonstrate near-perfect agreement with the proposed SPA-based expressions for both identical and non-identical amplitude response models, even under small RIS sizes. Furthermore, by applying an asymptotic SNR analysis to the SPA-based SER formulation, we reveal a theoretical relationship, where the coding gain is inversely proportional to the $N_t$-th negative moment of the non-zero eigenvalue of the RIS–receiver Gram matrix $\mathbf{\Phi^{\dagger} \mathbf{g}^{\dagger} \mathbf{g} \Phi}$. This insight enables a novel optimization strategy that eliminates the need for computationally intensive performance metrics such as SER, outage probability, or ergodic capacity. To demonstrate the practical applicability of the proposed objective function, we designed a lightweight optimization algorithm based on SAP and Greedy Search with RIS element grouping. When the grouping is chosen appropriately with respect to the number of RIS elements, the system effectively converges to the ideal amplitude response case.
\IEEEpubidadjcol

The primary focus of this work is to introduce a theoretically grounded objective function for RIS phase shift optimization. The implemented optimization is based on a Greedy Search strategy, which is effective when the group size is small but may become time-consuming as the number of groups increases. In real-world systems, this challenge can be addressed by using the proposed objective function to guide the training of AI-based models, enabling near-optimal phase configurations to be obtained with low processing time. Exploring such AI-driven solutions is left for future work. It is also important to highlight that this work adopts OSTBC to exploit diversity gains in RIS-assisted multiple antenna systems. As a result, the system operates in a virtual MISO mode, where the transmitter and receiver dynamically select antennas that comply with the OSTBC structure. This implicitly requires the use of transmit or receive antenna selection (TAS/RAS) mechanisms to maintain compatibility with the coding scheme. In order to fully exploit spatial multiplexing gains and address this structural limitation, future work will explore generalized space-time coding strategies or alternative diversity techniques designed for RIS-assisted MIMO architectures.

\section*{APPENDIX}
\subsection{LCLT Derivation for Theorem 4}
\label{appendix:proof_gaussian}
\IEEEpubidadjcol
\renewcommand{\theequation}{A.\arabic{equation}}
\setcounter{equation}{0}

This appendix outlines the justification for approximating the sum \( \lambda = \sum_{i=1}^{N_{\text{RIS}}} X_i \), where \( X_i \sim \text{Exp}\left( \frac{1}{\beta_i^2(\phi_i)} \right) \), each with mean \( \mu_i = \beta_i^2(\phi_i) \) and variance \( \sigma_i^2 = \beta_i^4(\phi_i) \). These variables are mutually independent and non-identically distributed. Let the mean and variance of each summand be
\begin{subequations}
\begin{equation}
\mu_i = \mathbb{E}[X_i] = \beta_i^2(\phi_i), 
\end{equation}
\begin{equation}
\sigma_i^2 = \mathrm{Var}(X_i) = \beta_i^4(\phi_i),
\end{equation}
\end{subequations}
and define the normalized sum as
\begin{equation}
Z_N = \frac{1}{s_N} \sum_{i=1}^{N_{\text{RIS}}} (X_i - \mu_i),
\end{equation}
where
 \[ 
 s_N^2 = \sum_{i=1}^{N_{\text{RIS}}} \sigma_i^2.
\]
To invoke the Lindeberg-Feller CLT we consider the Lindeberg condition, which requires that for every \( \varepsilon > 0 \),
\begin{equation}
\label{eqn:lindeberg}
\lim_{N_{\text{RIS}} \to \infty} \frac{1}{s_N^2} \sum_{i=1}^{N_{\text{RIS}}}
\mathbb{E} \left[ (X_i - \mu_i)^2 \cdot \mathbf{1}_{\{|X_i - \mu_i| > \varepsilon s_N\}} \right] = 0.
\end{equation}
where \(\mathbf{1}_{\{\cdot\}}\) denotes the indicator operator.
In practical settings, where the scale parameters \( \beta_i^2(\phi_i) \) are bounded and finite, the exponential tail of \( X_i \) ensures that the probability of large deviations \( |X_i - \mu_i| > \varepsilon s_N \) decays rapidly. Consequently, the contribution of such terms to the summation vanishes as \( N_{\text{RIS}} \to \infty \). Indeed, under the assumption that no single variance dominates the total variance, i.e.,
\begin{equation}
\label{eqn:variance_ratio}
\max_{1 \leq i \leq N_{\text{RIS}}} \frac{\sigma_i^2}{s_N^2} \to 0,
\end{equation}
as \( N_{\text{RIS}} \to \infty \), the contribution of any individual sum becomes asymptotically negligible in the normalized sum. This ensures that \( 1/N_{\text{RIS}} \to 0 \) implies an asymptotically vanishing influence of outlier terms in~\eqref{eqn:lindeberg}, effectively satisfying the Lindeberg condition.
Hence, by the Lindeberg-Feller CLT, the normalized sum converges in distribution to the standard normal distribution \cite{G-CLT}
\begin{equation}
Z_N \overset{d}{\to} \mathcal{N}(0,1) \quad \text{as} \quad N_{\text{RIS}} \to \infty.
\end{equation}
Therefore, the original sum \( \lambda \) asymptotically follows a Gaussian distribution:
\begin{equation}
\lambda \sim \mathcal{N}(\mu_{\lambda}, \sigma_{\lambda}^2),
\end{equation}
where
\begin{subequations}
\begin{equation}
\mu_{\lambda} = \sum_{i=1}^{N_{\text{RIS}}} \mu_i = \sum_{i=1}^{N_{\text{RIS}}} \beta_i^2(\phi_i), 
\end{equation}
\begin{equation}
\sigma_{\lambda}^2 = \sum_{i=1}^{N_{\text{RIS}}} \sigma_i^2 = \sum_{i=1}^{N_{\text{RIS}}} \beta_i^4(\phi_i).
\end{equation}
\end{subequations}
This concludes the derivation.

\subsection{SPA Derivation for Theorem 5}
\label{appendix:proof_saddle}
\IEEEpubidadjcol
\renewcommand{\theequation}{B.\arabic{equation}}
\setcounter{equation}{0}

This appendix derives the SPA of the PDF \( f_{\lambda}(y) \) for \( \lambda = \sum_{i=1}^{N_{\text{RIS}}} X_i \), where \( X_i \sim \text{Exp}\left( \frac{1}{\beta_i^2(\phi_i)} \right) \) are mutually independent with distinct parameters.

The MGF of each \( X_i \) is given by \cite{shankar}
\begin{equation}
\label{App_B1}
M_{X_i}(s) = \left( 1 - s \beta^2_i(\phi_i) \right)^{-1}, \quad s < \frac{1}{\beta^2_i(\phi_i)}.
\end{equation}
Hence, the MGF of \( \lambda \) is
\begin{equation}
\label{App_B2}
M_{\lambda}(s) = \prod_{i=1}^{N_{\text{RIS}}} \left( 1 - s \beta^2_i(\phi_i) \right)^{-1}.
\end{equation}
Taking the natural logarithm yields the CGF \cite{daniels1954saddlepoint}
\begin{equation}
\label{App_B3}
\psi(s) = \log M_{\lambda}(s) = -\sum_{i=1}^{N_{\text{RIS}}} \log(1 - s \beta^2_i(\phi_i)).
\end{equation}
Differentiating \( \psi(s) \), we obtain:
\begin{subequations}
\begin{align}
\label{App_B4a}
\psi'(s) &= \sum_{i=1}^{N_{\text{RIS}}} \frac{\beta^2_i(\phi_i)}{1 - s \beta^2_i(\phi_i)}, \\
\label{App_B4b}
\psi''(s) &= \sum_{i=1}^{N_{\text{RIS}}} \frac{\beta^4_i(\phi_i)}{\left( 1 - s \beta^2_i(\phi_i) \right)^2}.
\end{align}
\end{subequations}
The saddle point \( \hat{s} \) is defined as the solution of
\begin{equation}
\label{App_B5}
\psi'(\hat{s}) = y.
\end{equation}
Once \( \hat{s} \) is obtained for a given \( y \), the SPA approximation of the PDF is computed by substituting \( \hat{s} \) into the expression:
\begin{equation}
\label{App_B6}
f_{\lambda}(y) \approx \frac{1}{\sqrt{2\pi \psi''(\hat{s})}} \exp\left( \psi(\hat{s}) - \hat{s}y \right).
\end{equation}
This concludes the derivation.

\section*{ACKNOWLEDGMENT}
This work was supported by the Communications and Signal Processing Research (H{\.{I}}SAR) Laboratory, T{\"{U}}B{\.{I}}TAK-B{\.{I}}LGEM, T{\"{U}}RK{\.{I}}YE. The authors would like to thank all the H{\.{I}}SAR researchers for their insightful and inspiring comments.

\bibliographystyle{IEEEtran}
\bibliography{ref}

\end{document}